\documentclass[12pt,english]{article}
\usepackage{mathptmx}
\usepackage[T1]{fontenc}
\usepackage[latin9]{inputenc}
\usepackage{geometry}
\usepackage{color}
\usepackage{babel}
\usepackage{mathtools}
\usepackage{amsmath}
\usepackage{amssymb}
\usepackage{stmaryrd}
\usepackage{graphicx}
\usepackage{setspace}
\usepackage[authoryear]{natbib}
\makeatletter
\@ifundefined{date}{}{\date{}}

\usepackage{pdfsync}
\usepackage[font={small}]{caption}

\bibpunct{(}{)}{,}{a}{,}{,}

\usepackage{amsmath}
\usepackage{subfig}
\usepackage{floatrow}
\usepackage{dsfont}

\usepackage{newtxmath}
\usepackage{newtxtext}
\usepackage{graphicx}

\makeatother

\begin{document}
\global\long\def\real{\mathbb{R}}%
 
\global\long\def\RefVol{\Omega_{0}}%
 
\global\long\def\Refx{\mathbf{X}}%
 
\global\long\def\Curx{\mathbf{x}}%
 
\global\long\def\map{\boldsymbol{\chi}}%
 
\global\long\def\defgrad{\mathbf{F}}%
 
\global\long\def\defgradT{\mathbf{F}^{\mathrm{T}}}%
 
\global\long\def\defgradTi{\mathbf{F}^{\mathrm{-T}}}%
 
\global\long\def\d{\mathrm{d}}%
 
\global\long\def\RCG{\mathbf{C}}%
 
\global\long\def\LCG{\mathbf{b}}%
 
\global\long\def\div#1{\nabla\cdot#1}%
 
\global\long\def\curl#1{\nabla\times#1}%
 
\global\long\def\T#1{#1^{\mathrm{T}}}%
 
\global\long\def\CurStress{\boldsymbol{\sigma}}%
 
\global\long\def\Refcurl#1{\nabla_{\mathbf{X}}\times#1}%
 
\global\long\def\s#1{#1^{\star}}%
 
\global\long\def\m#1{#1^{(m)}}%
 
\global\long\def\f#1{#1^{(f)}}%
 
\global\long\def\p#1{#1^{(p)}}%
 
\global\long\def\Refdiv#1{\nabla_{\mathbf{X}}\cdot#1}%
 
\global\long\def\phase#1{#1^{\left(p\right)}}%
 
\global\long\def\jm{j_{m}^{(p)}}%
 
\global\long\def\xinc{\mathbf{\dot{x}}}%
 
\global\long\def\Pinc{\mathbf{\dot{P}}}%
 
\global\long\def\Einc{\mathbf{\dot{E}}}%
 
\global\long\def\Finc{\mathbf{\dot{F}}}%
 
\global\long\def\dinc{\mathbf{\check{d}}}%
 
\global\long\def\einc{\mathbf{\check{e}}}%
 
\global\long\def\Dinc{\mathbf{\dot{D}}}%
 
\global\long\def\sinc{\boldsymbol{\Sigma}}%
 
\global\long\def\fA{\mathbb{\boldsymbol{\mathscr{A}}}}%
 
\global\long\def\fB{\mathbb{\boldsymbol{\mathscr{B}}}}%
 
\global\long\def\fC{\mathbb{\boldsymbol{\mathscr{C}}}}%
 
\global\long\def\plane{\left(x_{1},x_{3}\right)}%
 
\global\long\def\anti{\dot{x}_{2}}%
 
\global\long\def\G{\mathbf{G}}%
 
\global\long\def\Gt{\mathbf{G}'}%
 
\global\long\def\k{\mathbf{k}}%
 
\global\long\def\Gtk{\left(\Gt+\k\right)}%
 
\global\long\def\Gk{\left(\G+\k\right)}%
 
\global\long\def\GGtk{\left(\G+\Gt+\k\right)}%
 
\global\long\def\ebias{\hat{e}}%
 
\global\long\def\deformation{\boldsymbol{\chi}}%
 
\global\long\def\dg{\mathbf{F}}%
 
\global\long\def\dgcomp#1{F_{#1}}%
 
\global\long\def\piola{\mathbf{P}}%
 
\global\long\def\refbody{\Omega_{0}}%
 
\global\long\def\refbnd{\partial\refbody}%
 
\global\long\def\bnd{\partial\Omega}%
 
\global\long\def\rcg{\mathbf{C}}%
 
\global\long\def\lcg{\mathbf{b}}%
 
\global\long\def\rcgcomp#1{C_{#1}}%
 
\global\long\def\cronck#1{\delta_{#1}}%
 
\global\long\def\lcgcomp#1{b_{#1}}%
 
\global\long\def\deformation{\boldsymbol{\chi}}%
 
\global\long\def\dgt{\dg^{\mathrm{T}}}%
 
\global\long\def\idgcomp#1{F_{#1}^{-1}}%
 
\global\long\def\velocity{\mathbf{v}}%
 
\global\long\def\accel{\mathbf{a}}%
 
\global\long\def\vg{\mathbf{l}}%
 
\global\long\def\idg{\dg^{-1}}%
 
\global\long\def\cauchycomp#1{\sigma_{#1}}%
 
\global\long\def\idgt{\dg^{\mathrm{-T}}}%
 
\global\long\def\cauchy{\boldsymbol{\sigma}}%
 
\global\long\def\normal{\mathbf{n}}%
 
\global\long\def\normall{\mathbf{N}}%
 
\global\long\def\traction{\mathbf{t}}%
 
\global\long\def\tractionl{\mathbf{t}_{L}}%
 
\global\long\def\ed{\mathbf{d}}%
 
\global\long\def\edcomp#1{d_{#1}}%
 
\global\long\def\edl{\mathbf{D}}%
 
\global\long\def\edlcomp#1{D_{#1}}%
 
\global\long\def\ef{\mathbf{e}}%
 
\global\long\def\efcomp#1{e_{#1}}%
 
\global\long\def\efl{\mathbf{E}}%
 
\global\long\def\freech{q_{e}}%
 
\global\long\def\surfacech{w_{e}}%
 
\global\long\def\outer#1{#1^{\star}}%
 
\global\long\def\perm{\epsilon_{0}}%
 
\global\long\def\matper{\epsilon}%
 
\global\long\def\jump#1{\llbracket#1\rrbracket}%
 
\global\long\def\identity{\mathbf{I}}%
 
\global\long\def\area{\mathrm{d}a}%
 
\global\long\def\areal{\mathrm{d}A}%
 
\global\long\def\refsys{\mathbf{X}}%
 
\global\long\def\Grad{\nabla_{\refsys}}%
 
\global\long\def\grad{\nabla}%
 
\global\long\def\divg{\nabla\cdot}%
 
\global\long\def\Div{\nabla_{\refsys}}%
 
\global\long\def\derivative#1#2{\frac{\partial#1}{\partial#2}}%
 
\global\long\def\aef{\Psi}%
 
\global\long\def\dltendl{\edl\otimes\edl}%
 
\global\long\def\ii#1{I_{#1}}%
 
\global\long\def\dh{\hat{D}}%
 
\global\long\def\inc#1{\dot{#1}}%
 
\global\long\def\sys{\mathbf{x}}%
 
\global\long\def\curl{\nabla}%
 
\global\long\def\Curl{\nabla_{\refsys}}%
 
\global\long\def\piolaincpush{\boldsymbol{\Sigma}}%
 
\global\long\def\piolaincpushcomp#1{\Sigma_{#1}}%
 
\global\long\def\edlincpush{\check{\mathbf{d}}}%
 
\global\long\def\edlincpushcomp#1{\check{d}_{#1}}%
 
\global\long\def\efincpush{\check{\mathbf{e}}}%
 
\global\long\def\efincpushcomp#1{\check{e}_{#1}}%
 
\global\long\def\elaspush{\boldsymbol{\mathcal{C}}}%
 
\global\long\def\elecpush{\boldsymbol{\mathcal{A}}}%
 
\global\long\def\elaselecpush{\boldsymbol{\mathcal{B}}}%
 
\global\long\def\disgrad{\mathbf{h}}%
 
\global\long\def\disgradcomp#1{h_{#1}}%
 
\global\long\def\trans#1{#1^{\mathrm{T}}}%
 
\global\long\def\phase#1{#1^{\left(p\right)}}%
 
\global\long\def\elecpushcomp#1{\mathcal{A}_{#1}}%
 
\global\long\def\elaselecpushcomp#1{\mathcal{B}_{#1}}%
 
\global\long\def\elaspushcomp#1{\mathcal{C}_{#1}}%
 
\global\long\def\dnh{\aef_{DH}}%
 
\global\long\def\woo{\varsigma}%
 
\global\long\def\wif{\Lambda}%
 
\global\long\def\structurefun{S}%
 
\global\long\def\dg{\mathbf{F}}%
 
\global\long\def\dgcomp#1{F_{#1}}%
 
\global\long\def\piola{\mathbf{P}}%
 
\global\long\def\refbody{\Omega_{0}}%
 
\global\long\def\refbnd{\partial\refbody}%
 
\global\long\def\bnd{\partial\Omega}%
 
\global\long\def\rcg{\mathbf{C}}%
 
\global\long\def\lcg{\mathbf{b}}%
 
\global\long\def\rcgcomp#1{C_{#1}}%
 
\global\long\def\cronck#1{\delta_{#1}}%
 
\global\long\def\lcgcomp#1{b_{#1}}%
 
\global\long\def\deformation{\boldsymbol{\chi}}%
 
\global\long\def\dgt{\dg^{\mathrm{T}}}%
 
\global\long\def\idgcomp#1{F_{#1}^{-1}}%
 
\global\long\def\velocity{\mathbf{v}}%
 
\global\long\def\accel{\mathbf{a}}%
 
\global\long\def\vg{\mathbf{l}}%
 
\global\long\def\idg{\dg^{-1}}%
 
\global\long\def\cauchycomp#1{\sigma_{#1}}%
 
\global\long\def\idgt{\dg^{\mathrm{-T}}}%
 
\global\long\def\cauchy{\boldsymbol{\sigma}}%
 
\global\long\def\normal{\mathbf{n}}%
 
\global\long\def\normall{\mathbf{N}}%
 
\global\long\def\traction{\mathbf{t}}%
 
\global\long\def\tractionl{\mathbf{t}_{L}}%
 
\global\long\def\ed{\mathbf{d}}%
 
\global\long\def\edcomp#1{d_{#1}}%
 
\global\long\def\edl{\mathbf{D}}%
 
\global\long\def\edlcomp#1{D_{#1}}%
 
\global\long\def\ef{\mathbf{e}}%
 
\global\long\def\efcomp#1{e_{#1}}%
 
\global\long\def\efl{\mathbf{E}}%
 
\global\long\def\freech{q_{e}}%
 
\global\long\def\surfacech{w_{e}}%
 
\global\long\def\outer#1{#1^{\star}}%
 
\global\long\def\perm{\epsilon_{0}}%
 
\global\long\def\matper{\epsilon}%
 
\global\long\def\jump#1{\llbracket#1\rrbracket}%
 
\global\long\def\identity{\mathbf{I}}%
 
\global\long\def\area{\mathrm{d}a}%
 
\global\long\def\areal{\mathrm{d}A}%
 
\global\long\def\refsys{\mathbf{X}}%
 
\global\long\def\Grad{\nabla_{\refsys}}%
 
\global\long\def\grad{\nabla}%
 
\global\long\def\divg{\nabla\cdot}%
 
\global\long\def\Div{\nabla_{\refsys}}%
 
\global\long\def\derivative#1#2{\frac{\partial#1}{\partial#2}}%
 
\global\long\def\aef{\Psi}%
 
\global\long\def\dltendl{\edl\otimes\edl}%
 
\global\long\def\tr#1{\mathrm{tr}\left(#1\right)}%
 
\global\long\def\dh{\hat{D}}%
 
\global\long\def\lage{\mathbf{E}}%
 
\global\long\def\inc#1{\dot{#1}}%
 
\global\long\def\sys{\mathbf{x}}%
 
\global\long\def\curl{\nabla}%
 
\global\long\def\Curl{\nabla_{\refsys}}%
 
\global\long\def\piolaincpush{\boldsymbol{\Sigma}}%
 
\global\long\def\piolaincpushcomp#1{\Sigma_{#1}}%
 
\global\long\def\edlincpush{\check{\mathbf{d}}}%
 
\global\long\def\edlincpushcomp#1{\check{d}_{#1}}%
 
\global\long\def\efincpush{\check{\mathbf{e}}}%
 
\global\long\def\efincpushcomp#1{\check{e}_{#1}}%
 
\global\long\def\elaspush{\boldsymbol{\mathcal{C}}}%
 
\global\long\def\elecpush{\boldsymbol{\mathcal{A}}}%
 
\global\long\def\elaselecpush{\boldsymbol{\mathcal{B}}}%
 
\global\long\def\disgrad{\mathbf{h}}%
 
\global\long\def\disgradcomp#1{h_{#1}}%
 
\global\long\def\trans#1{#1^{\mathrm{T}}}%
 
\global\long\def\phase#1{#1^{\left(p\right)}}%
 
\global\long\def\elecpushcomp#1{\mathcal{A}_{#1}}%
 
\global\long\def\elaselecpushcomp#1{\mathcal{B}_{#1}}%
 
\global\long\def\elaspushcomp#1{\mathcal{C}_{#1}}%
 
\global\long\def\dnh{\aef_{DG}}%
 
\global\long\def\dnhc{\mu\lambda^{2}}%
 
\global\long\def\dnhcc{\frac{\mu}{\lambda^{2}}+\frac{1}{\matper}d_{2}^{2}}%
 
\global\long\def\dnhb{\frac{1}{\matper}d_{2}}%
 
\global\long\def\afreq{\omega}%
 
\global\long\def\dispot{\phi}%
 
\global\long\def\edpot{\varphi}%
 
\global\long\def\kh{\hat{k}}%
 
\global\long\def\afreqh{\hat{\afreq}}%
 
\global\long\def\phasespeed{c}%
 
\global\long\def\bulkspeed{c_{B}}%
 
\global\long\def\speedh{\hat{c}}%
 
\global\long\def\dhth{\dh_{th}}%
 
\global\long\def\bulkspeedl{\bulkspeed_{\lambda}}%
 
\global\long\def\khth{\hat{k}_{th}}%
 
\global\long\def\p#1{#1^{\left(p\right)}}%
 
\global\long\def\maxinccomp#1{\inc{\outer{\sigma}}_{#1}}%
 
\global\long\def\maxcomp#1{\outer{\sigma}_{#1}}%
 
\global\long\def\relper{\matper_{r}}%
 
\global\long\def\sdh{\hat{d}}%
 
\global\long\def\iee{\varphi}%
 
\global\long\def\effectivemu{\tilde{\mu}}%
 
\global\long\def\fb#1{#1^{\left(a\right)}}%
 
\global\long\def\mt#1{#1^{\left(b\right)}}%
 
\global\long\def\phs#1{#1^{\left(p\right)}}%
 
\global\long\def\thc{h}%
 
\global\long\def\state{\mathbf{s}}%
 
\global\long\def\harmonicper{\breve{\matper}}%
 
\global\long\def\kb{k_{B}}%
 
\global\long\def\cb{\bar{c}}%
 
\global\long\def\mb{\bar{\mu}}%
 
\global\long\def\rb{\bar{\rho}}%
 
\global\long\def\wavenumber{k}%
 
\global\long\def\nh{\mathbf{n}}%
 
\global\long\def\mh{\mathbf{m}}%
 
\global\long\def\deflect{\inc x_{2}}%
 
\global\long\def\sdd#1{#1_{2,11}}%
 
\global\long\def\sdddd#1{#1_{2,1111}}%
 
\global\long\def\sd#1{#1_{2,1}}%
 
\global\long\def\sddd#1{#1_{2,111}}%
 
\global\long\def\xdddd#1{#1_{,\xi\xi\xi\xi}}%
 
\global\long\def\xdd#1{#1_{,\xi\xi}}%
 
\global\long\def\xd#1{#1_{,\xi}}%
 
\global\long\def\xddd#1{#1_{,\xi\xi\xi}}%
 
\global\long\def\jm{J_{m}}%
 
\global\long\def\dv{\Delta V}%
 
\global\long\def\ih{\mathbf{i}_{1}}%
 
\global\long\def\kh{\mathbf{i}_{3}}%
 
\global\long\def\jh{\mathbf{i}_{2}}%
 
\global\long\def\etil{E}%
 
\global\long\def\genT{\mathsf{Q}}%
 
\global\long\def\transfer{\mathsf{T}}%
 
\global\long\def\statevec{\mathbf{s}}%
 
\global\long\def\coefvec{\mathbf{c}}%
 
\global\long\def\pressure{p_{0}}%
 
\global\long\def\ncell#1{#1_{\left(n\right)}}%
 
\global\long\def\ydisp{\inc x_{2}}%
 
\global\long\def\ycord{x_{2}}%
 
\global\long\def\pn#1{\ncell{#1}^{\left(p\right)}}%
 
\global\long\def\pnm#1{#1_{\left(n\right)m}^{\left(p\right)}}%
 
\global\long\def\eigen{\alpha}%
 
\global\long\def\xcomp{x_{1}}%
 
\global\long\def\totalT{\mathsf{T_{\mathrm{tot}}}}%
 
\global\long\def\rads{\frac{\mathrm{rad}}{\mathrm{s}}}%
 
\global\long\def\lf{\gamma}%
 
\global\long\def\tf{T_{m}}%
 
\global\long\def\eigenim{\beta}%
 
\global\long\def\bS{\mathsf{S}}%
\global\long\def\dis#1{u_{#1}}%
\global\long\def\refden{\rho_{L}}%
\global\long\def\curden{\rho}%
 
\global\long\def\jump#1{\llbracket#1\rrbracket}%
 
\global\long\def\identity{\mathbf{I}}%
 
\global\long\def\area{\mathrm{d}a}%
 
\global\long\def\areal{\mathrm{d}A}%
 
\global\long\def\refsys{\mathbf{X}}%
 
\global\long\def\Grad{\nabla_{\refsys}}%
 
\global\long\def\grad{\nabla}%
 
\global\long\def\divg{\nabla\cdot}%
 
\global\long\def\Div{\nabla_{\refsys}}%
 
\global\long\def\derivative#1#2{\frac{\partial#1}{\partial#2}}%
 
\global\long\def\aef{\Psi}%
 
\global\long\def\dltendl{\edl\otimes\edl}%
 
\global\long\def\tr#1{\mathrm{tr}#1}%
 
\global\long\def\ii#1{I_{#1}}%
 
\global\long\def\dh{\hat{D}}%
 
\global\long\def\inc#1{\dot{#1}}%
 
\global\long\def\sys{\mathbf{x}}%
 
\global\long\def\curl{\nabla}%
 
\global\long\def\Curl{\nabla_{\refsys}}%
 
\global\long\def\piolaincpush{\boldsymbol{\Sigma}}%
 
\global\long\def\piolaincpushcomp#1{\Sigma_{#1}}%
 
\global\long\def\edlincpush{\check{\mathbf{d}}}%
 
\global\long\def\edlincpushcomp#1{\check{d}_{#1}}%
 
\global\long\def\efincpush{\check{\mathbf{e}}}%
 
\global\long\def\efincpushcomp#1{\check{e}_{#1}}%
 
\global\long\def\elaspush{\boldsymbol{\mathcal{C}}}%
 
\global\long\def\elecpush{\boldsymbol{\mathcal{A}}}%
 
\global\long\def\elaselecpush{\boldsymbol{\mathcal{B}}}%
 
\global\long\def\disgrad{\mathbf{h}}%
 
\global\long\def\disgradcomp#1{h_{#1}}%
 
\global\long\def\trans#1{#1^{\mathrm{T}}}%
 
\global\long\def\phase#1{#1^{\left(p\right)}}%
 
\global\long\def\elecpushcomp#1{\mathcal{A}_{#1}}%
\global\long\def\elaselecpushcomp#1{\mathcal{B}_{#1}}%
\global\long\def\elaspushcomp#1{\mathcal{C}_{#1}}%
 
\global\long\def\dnh{\aef_{DH}}%
\global\long\def\woo{\varsigma}%
\global\long\def\wif{\Lambda}%
\global\long\def\structurefun{S}%
\global\long\def\bondechden{\rho_{b}}%
\global\long\def\surfacebondech{w_{b}}%
\global\long\def\refbondechden{\rho_{B}}%
\global\long\def\refsurfacebondech{w_{B}}%
\global\long\def\incbondechden{\check{\rho}_{b}}%
 
\global\long\def\GmGt{\left(\G-\Gt\right)}%
\global\long\def\woo{\varsigma}%
\global\long\def\fA{\mathbb{\boldsymbol{\mathscr{A}}}}%
\global\long\def\fB{\mathbb{\boldsymbol{\mathscr{B}}}}%
\global\long\def\fC{\mathbb{\boldsymbol{\mathscr{C}}}}%
\global\long\def\hyper{\Psi}%

\global\long\def\piolacomp#1{P_{#1}}%
 
\global\long\def\derivativesec#1#2{\frac{\partial^{2}#1}{\partial#2^{2}}}%

\global\long\def\pressurestrain{\epsilon_{A}}%
\global\long\def\shearstrain{\epsilon_{T}}%
\global\long\def\pressurevel{v_{A}}%
\global\long\def\shearvel{v_{T}}%

\global\long\def\piolaonebyP{\alpha}%
\global\long\def\piolaonebyQ{\beta}%
\global\long\def\piolatwobyP{\gamma}%
\global\long\def\piolatwobyQ{\delta}%
\global\long\def\xbyt{c}%

\global\long\def\shearwavespeed{c_{-}}%
\global\long\def\pressurewavespeed{c_{+}}%

\global\long\def\piolapressure{\sigma}%
\global\long\def\piolashear{\tau}%

\global\long\def\ii#1{I_{#1}}%
\global\long\def\bulk{\kappa}%
\global\long\def\shear{\mu}%

\global\long\def\tr#1{\mathrm{tr}#1}%
 
\global\long\def\det#1{\mathrm{det}#1}%

\global\long\def\phaseA#1{#1^{\left(a\right)}}%
\global\long\def\phaseB#1{#1^{\left(b\right)}}%
\global\long\def\phase#1{#1^{\left(n\right)}}%

\global\long\def\state{\mathsf{s}}%
\global\long\def\flux{\mathsf{f}}%
\global\long\def\jump{a}%

\global\long\def\initial#1{#1^{\left(U\right)}}%
\global\long\def\amp#1{#1^{\left(L\right)}}%
\global\long\def\vc{d}%

\title{Oscillating vector solitary waves in soft laminates}
\author{Ron Ziv and Gal Shmuel\thanks{Corresponding author. Tel.: +1 972 778871613. \emph{E-mail address}:
meshmuel@technion.ac.il (G. Shmuel).}\\
 {\small{}{}{}Faculty of Mechanical Engineering, Technion\textendash Israel
Institute of Technology, Haifa 32000, Israel}\\
}
\maketitle
\begin{abstract}
Vector solitary waves are nonlinear waves of coupled polarizations
that propagate with constant velocity and shape. In mechanics, they
hold the potential to control locomotion, mitigate shocks and transfer
information, among other functionalities. Recently, such elastic waves
were numerically observed in compressible rubber-like laminates.
Here, we conduct numerical experiments to characterize the possible
vector solitary waves in these laminates,  and expose a new type
of waves whose amplitude and velocity oscillate periodically without
dispersing in time. This oscillation is a manifestation of a periodic
transfer of energy between the two wave polarizations, which we consider
as internal mode of the solitary wave. We find that the vector solitary
waves propagate faster at higher amplitudes, and determine a lower
bound for their velocity. We describe a procedure for identifying
which initial strains generate such vector solitary waves. This procedure
also enables an additional classification between tensile and compressive
solitary waves, according to the way that the axial strain changes
as the waves propagate.
\end{abstract}

\section{Introduction}

Nonlinearities in wave mechanics are the source of fascinating phenomena,
from instabilities \citep{Skipetrov200prl}, harmonic generation \citep{Saltiel2008prl,gonella17jmps,khajehtourian2019nonlinear},
extreme energy transfer \citep{ZHANG2018jmps} and non-reciprocal
transmission \citep{Lepri2011prl}, to the formation of shocks \citep{CHOCKALINGAM2020103746},
rogue waves \citep{Baronio2012prl} and \emph{solitary waves} \citep{SILLING2016jmps}.
The latter are amplitude-dependent waves which propagate with a constant
velocity and fixed shape, owing to a balance between dispersion and
nonlinearity in the system \citep{dauxois2006physics}. Initially
studied in fluid mechanics \citep{BOYD2015417}, quantum mechanics
\citep{Kasamatsu2006pra} and optics \citep{Stegeman1999science,kivshar2003optical},
solitary waves are recently gaining increased attention from the solid
mechanics community \citep{nadkarni2014dynamics,deng2018metamaterials,DENG2020jmps,Katz2019,mo2019cnoidal}.
This interest was triggered by the quest for mechanical \emph{metamaterials}\textemdash artificial
composites with properties and functionalities not found in nature
\citep{craster2012acoustic,Christensen2015MRCComunications,srivastava2015elastic,bertoldi2017flexible,Kadi2019nrp};
and its pertinent progress in additive manufacturing techniques for
creating these architectured materials \citep{raney2015}.

In addition to the mathematical significance of their analysis, the
study of solitary waves in mechanical systems shows also technological
potential in  impact mitigation \citep{yasuda2019origami}, mechanical
logic gates \citep{raney16}, nondestructive testing \citep{nasrollahi2017},
wave focusing \citep{deng2019focusing}, and robot locomotion \citep{Deng2020sciadv}.
The framework of these works is almost exclusively of discrete models,
with only few results on elastic \emph{continua}. A specific class
of continuum that exhibits nonlinearities is of soft materials, whose
microscopic composition \citep{arru&boyc93jmps} and ability to sustain
large deformations \citep{ogden97book} are the source of constitutive
and geometrical nonlinearities, respectively. Thus, soft materials
constitute a barely explored platform to theoretically study solitary
waves, and in turn experimentally realize using current 3D printing
capabilities \citep{Bandyopadhyay2015MRS,Truby2016,GARCIA2019am}.

To the best of our knowledge, the first numerical  observation of
one-dimensional solitary waves in elastic \emph{periodic} continua
was by \citet{leveque2002finite} and \citet{bale2002siam}, using
a finite-volume method they developed, which was followed by the studies
of \citet{yong2003solitary,andrianov2013dynamic,andrianov2014numerical,navarro2015nature,hussein2018nonlinear}
and the references therein. As explained in these works, the periodicity
of the media causes dispersion, whose balance with geometrical and
constitutive nonlinearities forms elastic solitary waves. Only recently,
this principle for generating solitary waves was tested by \citet{ziv2019b}
in \emph{two-dimensional} continuum elastodynamics. There, the authors
have developed a designated finite-volume method to simulate finite
motions of soft laminates whose displacement field consists of two
coupled components. By application of this method to a periodic repetition
of compressible \citet{gent96rc&t} layers that differ in their mass
density, the authors were able to observe the formation of \emph{vector}
solitary waves\textemdash solitary waves with (at least) two components
and two polarizations that are coupled one with the other\textemdash in
this case, when the axial and transverse components of the displacement
field are coupled. To the best of our knowledge, that was the first
report of vector solitary waves in an elastic continuum, within the
framework of nonlinear elastodynamics \citep{ogden97book}. An earlier
report of such waves in a \emph{discrete} mechanical medium was given
by \citet{deng2017,deng2019prl}, who conceived a model made of rigid
squares that are connected using linear springs at their corners,
thereby allowing for two coupled degrees of freedom for the squares:
rotation and translation.   While both the systems considered by
\citet{deng2017,deng2019prl} and the system considered here are two-dimensional
and nonlinear, there are fundamental differences between them, as
we explain in the sequel. These differences lead to opposite velocity-amplitude
relation for the vector solitary waves in each system.

Here, we carry out a comprehensive study of the vector solitary waves
that were observed by \citet{ziv2019b}, by analyzing a large set
of numerical experiments that were generated using the method they
developed in that work, with the following findings. First, we find
that these vector solitary waves can be divided into two types which
differ in their velocity and strain profile. We term the two types
\emph{quasi}-pressure and \emph{quasi}-shear solitary waves, since
in the limiting case when the heterogeneity vanishes and the strains
are small, they reduce to the standard pressure and shear waves, respectively.
We show that the profile of quasi-pressure vector solitary waves resembles
the sech$^{2}$ function, which is the solitary wave solution of the
well-known KdV equation \citep{korteweg1895xli}. We find that while
quasi-pressure solitary waves maintain an identical wave profile as
they pass between adjacent unit cells,  the profile and velocity
of quasi-shear solitary waves change between different cells in a
periodic and permanent manner. This oscillation occurs through a periodic
transfer of energy between the two polarizations of the vector solitary
wave, which we consider as internal mode of the solitary wave. Similar
long-lived oscillations were reported before in other fields \citep{Campbell1983ty,Mantsyzov1995pra,kivshar1998internal,Szankowski2010prl},
and were also reported recently in an acoustic system that is described
using the continuum limit of a discrete model based on mechanical\textendash electrical
analogies \citep{zhang2017bright}. Our results are the first report
of oscillating vector solitary waves that are generated from the equations
of continuum elastodynamics.

Next, we characterize the relation between the amplitude of the strain
 and the velocity of the solitary waves. This is done for a wide
range of initial conditions, which were carefully chosen to generate
in each experiment a single solitary wave. These experiments show
that the velocity is a monotonically increasing function of the amplitude.
This observation agrees with the results of \citet{ziv2019b}, who
generated trains of solitary waves from a single set of initial conditions,
and observed that the taller waves in the train propagate faster than
the shorter waves. Notably, this velocity-amplitude relation is opposite
to the relation for the vector solitary waves in the discrete mechanical
model of \citet{deng2017,deng2019prl}, there the waves with smaller
amplitudes are faster. We also determine the velocity of Bloch-Floquet
waves in the laminate in the limit of low-frequency, long-wavelength
linear elastodynamics \citep{santosa1991dispersive}, to find they
serve as a lower bound to the velocity of solitary waves in the nonlinear
settings. Our results thus generalize the findings of \citet{yong2003solitary}
and \citet{andrianov2014numerical}, who showed that one-dimensional
solitary waves in nonlinear laminates are supersonic\footnote{Solitary waves whose velocity exceeds the wave velocity in the linear
quasi-static limit are termed \emph{supersonic}.}, similarly to the feature of solitary wave solutions to the KdV equation
\citep{dauxois2006physics}.

Lastly, we describe a procedure for determining which initial deformations
will generate vector solitary waves. This is done using contour maps
of the characteristic velocities as functions of the strain, identifying
the initial strain and loading path in these maps, and employing a
certain criterion regarding the gradient of the characteristic velocity
along the path, as explained in detail later. These maps are also
useful in identifying how the axial strain in the laminate changes
during the propagation of the solitary waves. Specifically, we identify
two domains in the map for quasi-pressure solitary waves, namely,
a domain of solitary waves that tend to increase the axial strain
in the laminate and a separate domain of solitary waves that tend
to decrease the axial strain in the laminate. In accordance with this
tendency, we term the latter \emph{compressive} solitary waves, notwithstanding
the fact that the sign of axial strain in the laminate may be positive.
Similarly, we term the former \emph{tensile} solitary waves, notwithstanding
the fact that the sign of axial strain may be negative. Interestingly,
such a separation is absent from the map for quasi-shear solitary
waves, since these waves can only decrease the axial strain, as explained
later.

The rest of the paper is organized as follows. Sec.$\ $\ref{sec:Governing-equations}
contains a description of the relevant elastodynamics problem, together
with its governing vectorial equations. Sec.$\ $\ref{sec:Method-of-solution}
first revisits solutions of the differential relations between the
strain components in the benchmark problem of a homogeneous medium
\citep{ZIV2019mom}. In the second part of this Sec., the computational
study of the heterogeneous medium is carried out, where we also show
that solutions for the homogeneous medium serve as an estimators for
the differential relations in the main problem. A recap of our main
results together with comments on future work close the paper in Sec.$\ $\ref{sec:Summary}.

\section{Problem statement and governing equations\label{sec:Governing-equations}}

\begin{figure}[!t]
\sidesubfloat[]{\includegraphics[bb=0bp 0bp 518bp 602bp,scale=0.385]{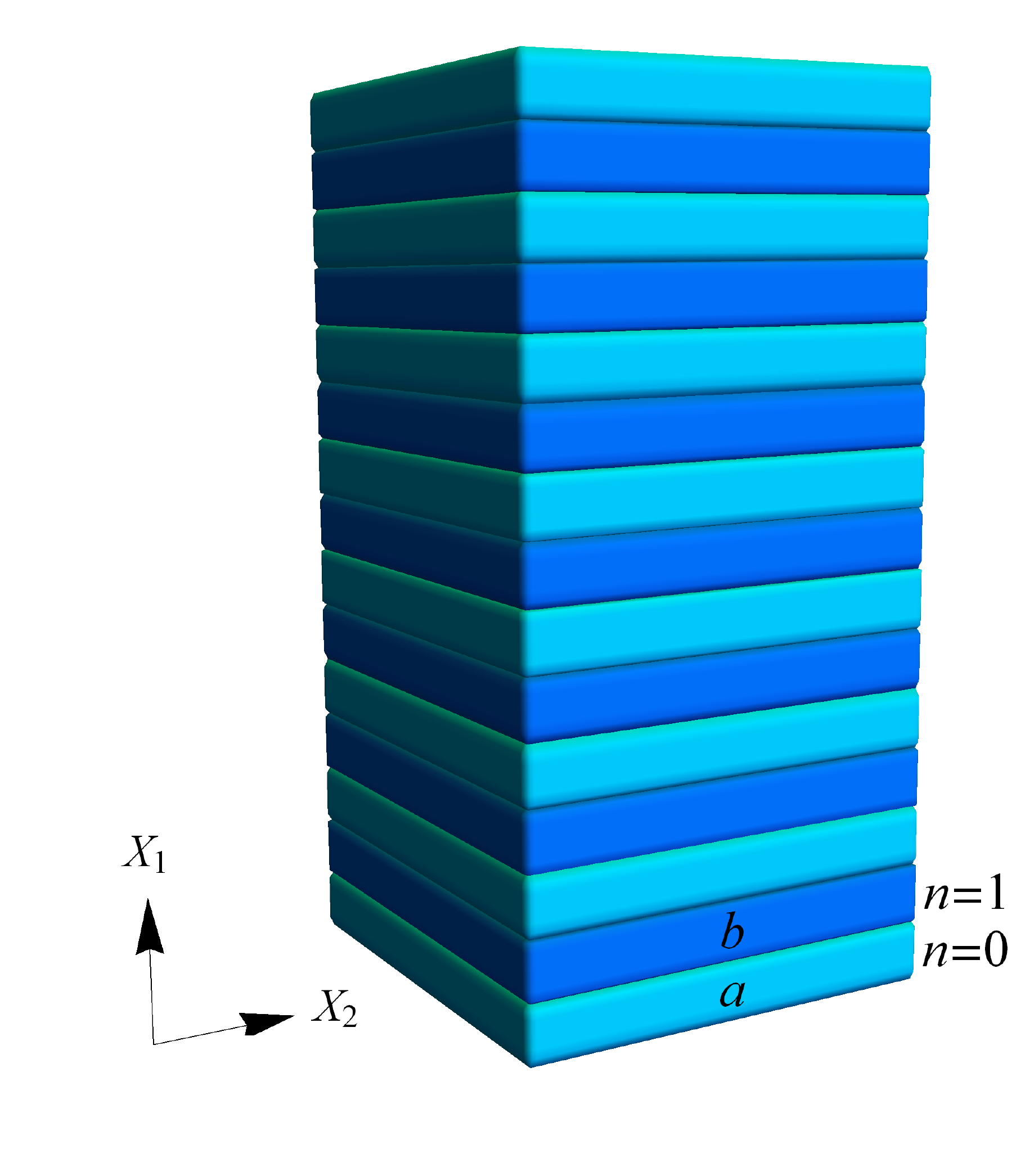}} \raisebox{-4\height}{{\LARGE{}$\overset{\map\left(t>0\right)}{\longrightarrow}$}}\sidesubfloat[]{\includegraphics[scale=0.3]{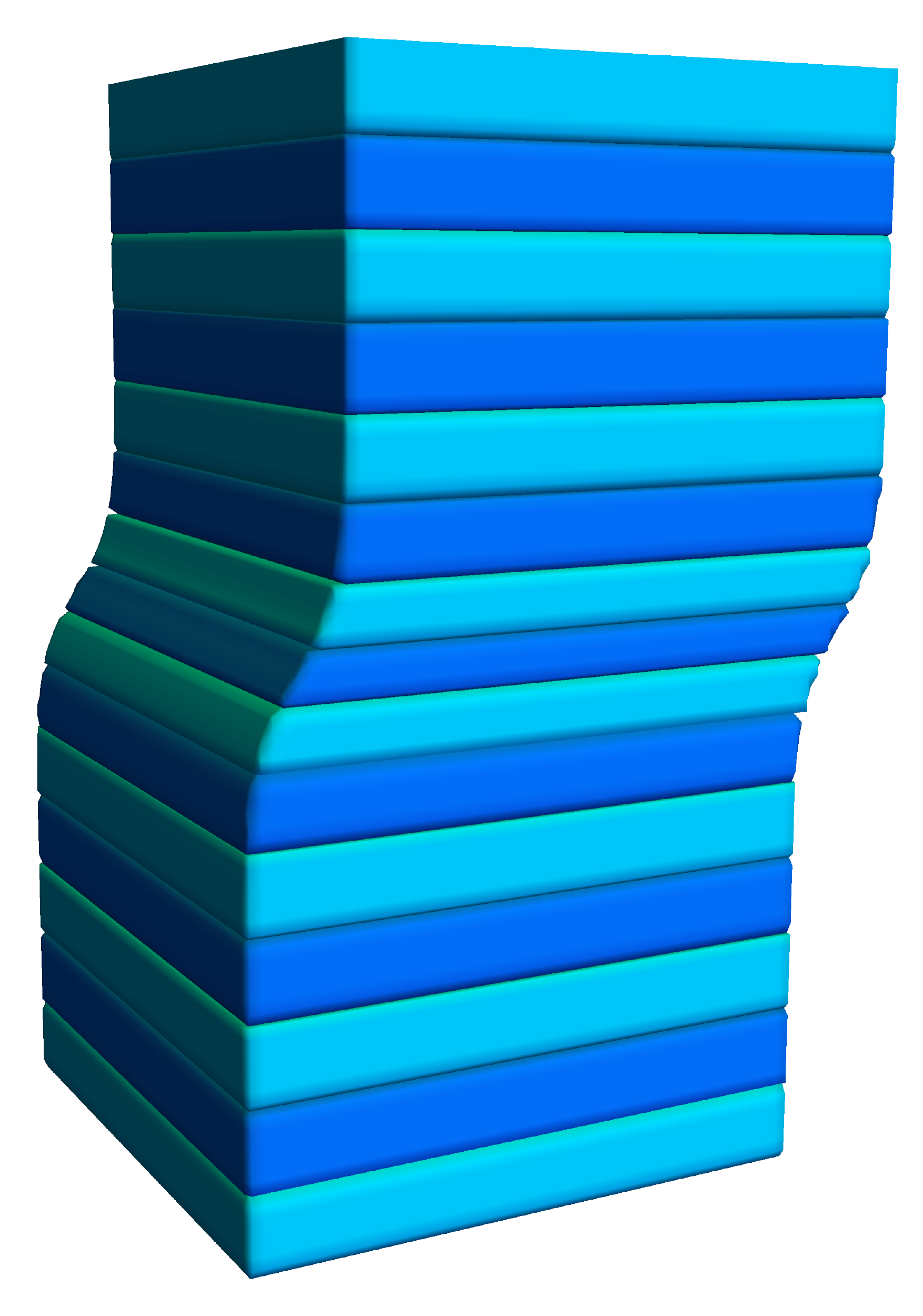}} 

\caption{Illustration of a two-phase periodic laminate (a) in the reference
configuration, and (b) undergoing a coupled axial and transverse deformation.\label{fig:Illustration}}
\end{figure}

We consider an infinite periodic repetition in the $X_{1}$ direction
of two hyperelastic phases, namely, $a$ and $b$, governed by the
same strain energy density function $\hyper$, and different initial
mass density $\refden$. We set $X_{1}=0$ at the beginning of a certain
$a$ phase, and denote its number by $n=0$, such that even and odd
values of $n$ correspond to layers made of phase $a$ and $b$, respectively.
At the initial state, the laminate is subjected to a combination of
axial and transverse displacement fields as functions of $X_{1}$.
Assuming that the laminate is uniform and infinite in $X_{2}$, the
resultant motion $\Curx=\map\left(\Refx,t\right)$ is defined by

\begin{equation}
x_{1}=X_{1}+\dis 1\left(X_{1},t\right),\;x_{2}=X_{2}+\dis 2\left(X_{1},t\right),\label{eq:deformation}
\end{equation}
and maps material points from the reference coordinate $\Refx$ to
the current coordinate $\Curx$ (Fig.$\ $\ref{fig:Illustration}).
The problem amounts to determining the displacements $u_{i}\left(X_{1},t>0\right)$
for a given initial state. This is carried out by solving the balance
equations for the corresponding first Piola-Kirchhoff stress tensor
$\piola=\grad_{\defgrad}\hyper$, where $\defgrad=\grad_{\text{\ensuremath{\Refx}}}\map$
is the deformation gradient. When restricting attention to smooth
waves\footnote{Roughly speaking, smooth waves propagate when the velocity at the
tail of the wave does not exceed the velocity at its front, and changes
monotonically in between these ends \citep{DAVISON1966249,ZIV2019mom,CHOCKALINGAM2020103746}.}, the balance equations are\floatsetup[figure]{style=plain,subcapbesideposition=top}
\begin{equation}
\Refdiv{\mathbf{\piola}}=\refden\deformation_{,tt},\label{eq:eom}
\end{equation}
where we recall that $\refden$ is a periodic function of $X_{1}$,
and it is tacitly assumed that $\map$ is continuous and twice differentiable,
except at the interfaces between phases. For the given mapping \eqref{eq:deformation},
Eq.$\ $\eqref{eq:eom} is reduced to

\begin{equation}
\begin{aligned}\piolacomp{11,1}=\refden\derivativesec{u_{1}}t,\ \piolacomp{21,1}=\refden\derivativesec{u_{2}}t.\end{aligned}
\label{eq:motion}
\end{equation}
In order to reformulate Eq.$\ $\eqref{eq:motion} in a more convenient
form for the forthcoming analysis, we first introduce the Lagrangian
strains and velocities of the axial and transverse displacements,
namely,
\begin{equation}
\pressurestrain:=\derivative{u_{1}}{X_{1}},\ \shearstrain:=\derivative{u_{2}}{X_{1}},\ \pressurevel:=\derivative{u_{1}}t,\ \shearvel:=\derivative{u_{2}}t.\label{eq:derivs}
\end{equation}
In the limiting case of a simple shear $\left(u_{1}=0\right)$, the
quantity $\shearstrain$ is called the amount of shear \citep{ogden97book},
where in the limiting case of a uniaxial deformation $\left(u_{2}=0\right)$,
the quantity $\pressurestrain$ is related to the axial stretch. We
use Eq.$\ $\eqref{eq:derivs} to carry out a reduction of order and
rewrite Eq.$\ $\eqref{eq:motion} as
\begin{equation}
\left(\begin{array}{c}
\pressurestrain\\
\shearstrain\\
\refden\pressurevel\\
\refden\shearvel
\end{array}\right)_{,t}+\left(\begin{array}{cccc}
0 & 0 & -\frac{1}{\refden} & 0\\
0 & 0 & 0 & -\frac{1}{\refden}\\
-\piolaonebyP & -\beta & 0 & 0\\
-\gamma & -\delta & 0 & 0
\end{array}\right)\left(\begin{array}{c}
\pressurestrain\\
\shearstrain\\
\refden\pressurevel\\
\refden\shearvel
\end{array}\right)_{,X_{1}}=\left(\begin{array}{c}
0\\
0\\
0\\
0
\end{array}\right),\label{eq:conservativeSystem}
\end{equation}
where

\begin{equation}
\piolaonebyP=\derivative{\piolacomp{11}}{\pressurestrain},\;\piolaonebyQ=\derivative{\piolacomp{11}}{\shearstrain},\;\piolatwobyP=\derivative{\piolacomp{21}}{\pressurestrain},\;\piolatwobyQ=\derivative{\piolacomp{21}}{\shearstrain}.\label{eq:defineCoeff}
\end{equation}
For hyperelastic materials we have that $\piolaonebyQ\equiv\piolatwobyP$
owing to the equality of mixed partials of $\hyper$.

The eigenvalues of Eq.$\ $\eqref{eq:conservativeSystem} are the
characteristic wave velocities at each material point, and are functions
of the phase and deformation at that point, given by

\begin{equation}
c=\pm\sqrt{\frac{1}{2\refden}\left[\piolaonebyP+\piolatwobyQ\pm\sqrt{\left(\piolaonebyP-\piolatwobyQ\right)^{2}+4\piolaonebyQ\piolatwobyP}\right]}\eqqcolon\pm c_{\pm},\label{eq:wavespeeds}
\end{equation}
where $c_{+}$ (resp.$\ $$c_{-}$) corresponds to the plus (resp.$\ $minus)
sign of the inner square root. In the limit of linear elasticity,
$c_{+}$ and $c_{-}$ are the velocities of pressure and shear waves,
respectively \citep{DAVISON1966249}. Since in finite elasticity the
corresponding modes involve both axial and transverse displacements,\emph{
i.e.}, there are no pure modes, we refer to those associated with
$\shearwavespeed$ and $\pressurewavespeed$ as \emph{quasi-}shear
and \emph{quasi-}pressure waves, respectively \citep{bland1965plane,ZIV2019mom}.

The eigenvectors of the matrix in Eq.$\ $\eqref{eq:conservativeSystem}
which are associated with the generated waves are

\begin{equation}
r_{1}=\left(\begin{array}{c}
1\\
\eta_{+}\\
\refden c_{+}\\
\eta_{+}\refden c_{+}
\end{array}\right),\;r_{2}=\left(\begin{array}{c}
-\eta_{-}\\
1\\
-\eta_{-}\refden c_{-}\\
\refden c_{-}
\end{array}\right),\ r_{3}=\left(\begin{array}{c}
-\eta_{-}\\
1\\
\eta_{-}\refden c_{-}\\
-\refden c_{-}
\end{array}\right),\;r_{4}=\left(\begin{array}{c}
1\\
\eta_{+}\\
-\refden c_{+}\\
-\eta_{+}\refden c_{+}
\end{array}\right),\label{eq:eigenvectors}
\end{equation}
where 
\begin{equation}
\eta_{+}=\frac{\refden c_{+}^{2}-\piolaonebyP}{\beta},\quad\eta_{-}=\frac{\beta}{\alpha-\refden c_{-}^{2}},\label{eq:eta}
\end{equation}
and the corresponding characteristic velocities are $-\pressurewavespeed,\,-\shearwavespeed,\,\shearwavespeed$
and $\pressurewavespeed$, respectively. Through algebraic manipulation
and using the fact that $\alpha,\beta,$ and $\delta$ are derived
from the same scalar potential, we find that $\text{\ensuremath{\eta_{+}}}$
is identical to $\eta_{-}$ and independent of $\rho_{L}$. The components
of the eigenvectors\textemdash and specifically $\eta_{\pm}$\textemdash are
differential relations between the strain and velocity fields. When
the initial conditions are piecewise-constant with a single jump discontinuity
and the medium is homogeneous, the problem is called a Riemann problem,
which can be solved by a variation of the method of characteristics.
The method exploits the fact that the solution can be expressed as
function of a single variable\textemdash here $X_{1}/t$\textemdash in
which case the corresponding waves are called simple waves \citep{davison2008book}.
The resultant equations can be solved using standard methods such
as Runge\textendash Kutta methods. \citet{ZIV2019mom} have adapted
such an approach to calculate smooth waves and shocks in semi-infinite
compressible Gent media, when subjected to combined transverse and
axial impacts. The challenge in solving Eq.$\ $\eqref{eq:conservativeSystem}
when the medium is heterogeneous and the initial data is not piecewise-uniform
is associated with the interactions between waves that are repeatedly
scattered at the interfaces. In this case, the ansatz of simple waves
fails and the previous approach is no longer applicable. This brings
a need for more sophisticated numerical solvers. The course taken
by \citet{ziv2019b} was to develop a designated algorithm based on
a finite-volume method\footnote{For an excellent treatise on finite-volume methods, the reader is
referred to the book of \citet{LeVeque2002book}.}, by adapting the schemes of \citet{LEVEQUE1997327,leveque2002finite}
and \citet{bale2002siam}, to the type of problem considered here;
this adaptation is used in the study described next.

\section{Computational study\label{sec:Method-of-solution}}

\emph{Homogeneous benchmark problem}.---Before we proceed to the
study of the main problem, it is advantageous to present the strain
relations of the homogeneous benchmark problem. As we will show in
the sequel, the reason is that they serve as estimators to the laminated
case. Thus, the differential relation between the axial and shear
strains in quasi-shear and quasi-pressure waves traversing homogeneous
media are given by \citep{DAVISON1966249,ZIV2019mom}

\begin{subequations}
\label{integralCurves}
\begin{equation}
\begin{aligned}\textrm{quasi-pressure:\ \ } & \derivative{\shearstrain}{\pressurestrain}=\eta_{+},\end{aligned}
\label{eq:integralCurvesPressure}
\end{equation}
\label{etea}
\begin{equation}
\begin{aligned}\textrm{\;\;\;\;\ \;\;\;\,quasi-shear:\ \ } & \derivative{\shearstrain}{\pressurestrain}=-\eta_{-}^{-1},\end{aligned}
\label{eq:integralCurvesShear}
\end{equation}
\end{subequations}
with the compatibility condition 
\begin{equation}
\shearstrain\left(\pressurestrain=\initial{\pressurestrain}\right)=\initial{\shearstrain},\label{eq:comp0}
\end{equation}
where $\initial{\pressurestrain}$ and $\initial{\shearstrain}$ can
be associated with axial and shear pre-strains, respectively. Eq.$\ $\eqref{etea}
is intimately related to the Poynting effect \citep{Poynting1909prsa},
namely, the axial deformation that accompanies finite shear, hence
occurs only at nonlinear deformations \citep{ogden97book}. Thus,
this effect is prominent in soft materials, which are capable of undergoing
large strains \citep{Cioroianu2013PRE,Horgan2017prsa}. The exact
manner in which this effect takes place depends on the constitutive
response of the material \citep{Mihai2017nr}. Accordingly, the solution
of Eq.$\ $\eqref{integralCurves} also depends on the constitutive
response of the material, which in our study is modeled using the
compressible \citet{gent96rc&t} model with the strain energy function
\citep{olp2007}
\begin{equation}
\hyper\left(\defgrad\right)=-\frac{\mu\jm}{2}\ln\left(1-\frac{\tr{\defgradT\defgrad}-3}{\jm}\right)-\mu\ln\det{\defgrad}+\left(\frac{\bulk}{2}-\frac{\mu}{3}-\frac{\mu}{\jm}\right)\left(\det{\defgrad}-1\right)^{2};\label{eq:Gent}
\end{equation}
here, $\mu$ and $\kappa$ coincide with the shear and bulk moduli
in the limit of small strains, respectively, and the dimensionless
parameter $\jm$ models a stiffening of strain. The Gent model was
originally developed to capture the stiffening of rubber materials
\citep{arru&boyc93jmps}. Recently, the model was also shown suitable
for capturing nonlinear wave phenomena such as shear shocks and tensile-induced
shocks, which were experimentally observed by \citet{Catheline2003PRL}
and \citet{espindola2017shear}, and \citet{NIEMCZURA2011442}, respectively.

\begin{figure}[!t]
\sidesubfloat[]{\includegraphics[scale=0.44]{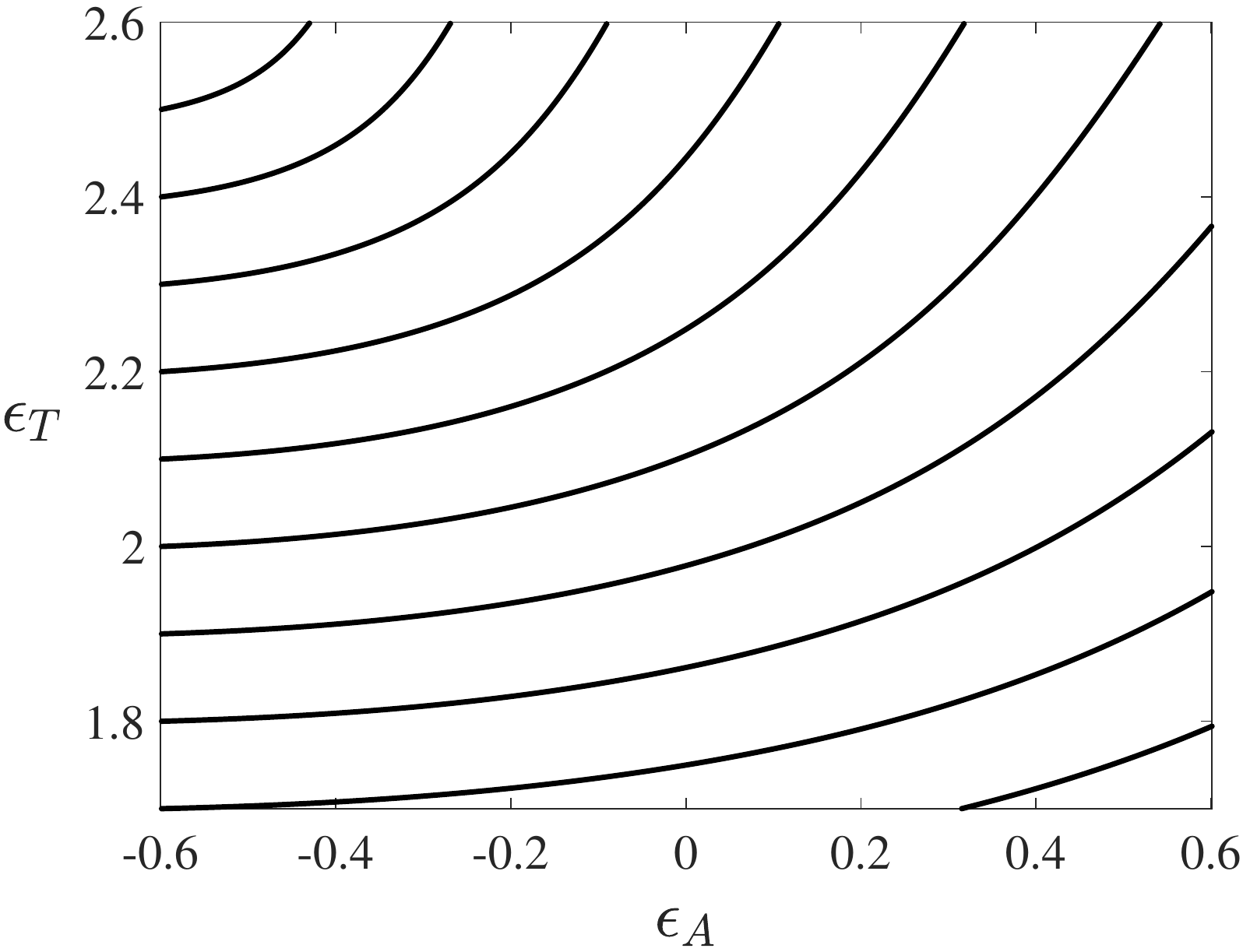}}  $\ \ \ \ \ $\sidesubfloat[]{\includegraphics[scale=0.44]{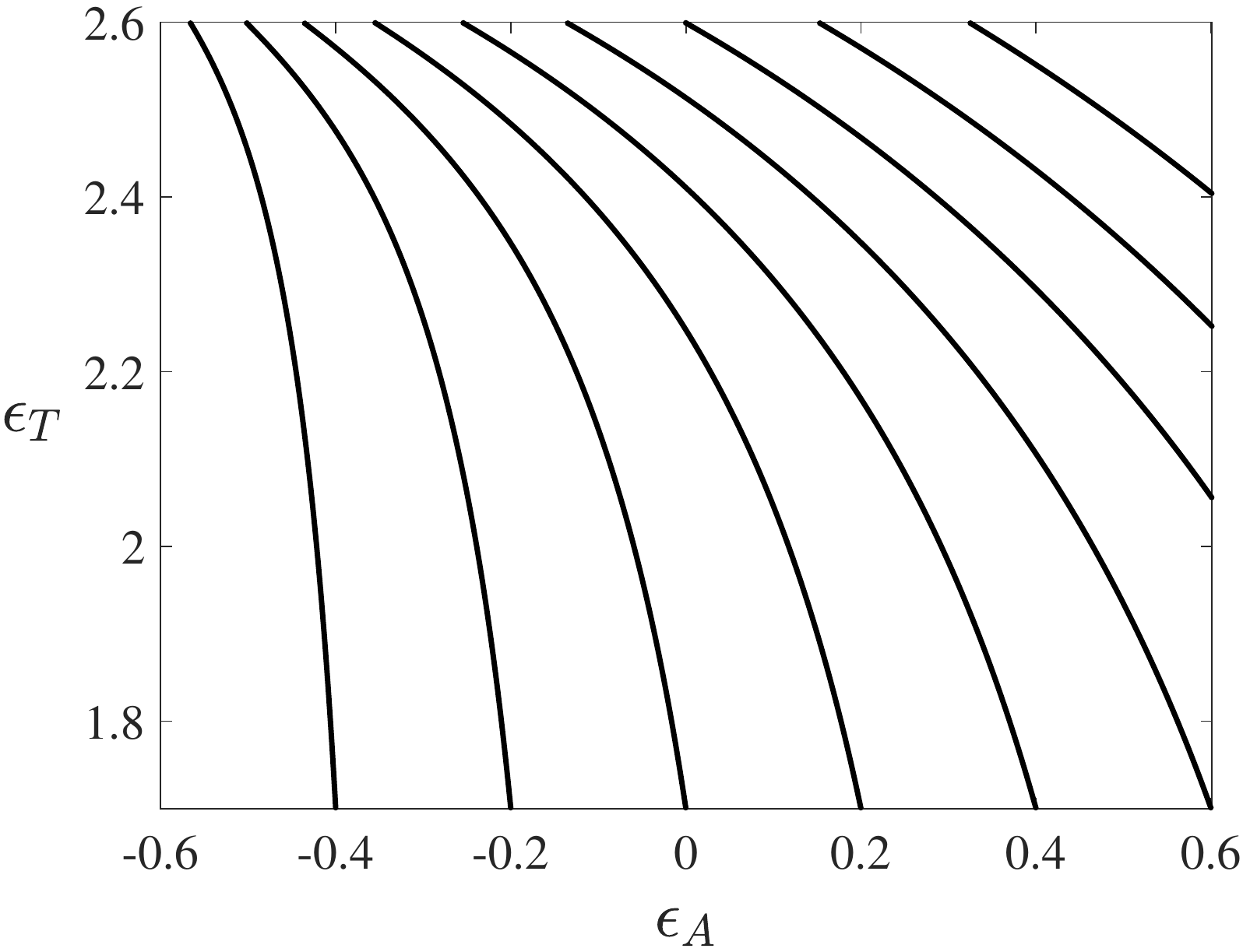}}  

\caption{Numerical solutions of Eq.$\ $\eqref{eq:integralCurvesPressure},
associated with quasi-pressure waves, and Eq.$\ $\eqref{eq:integralCurvesShear},
associated with quasi-shear waves, for a Gent material with the model
parameters \eqref{eq:paremeters} and different values of $\protect\initial{\protect\pressurestrain}$
and $\protect\initial{\protect\shearstrain}$.\label{fig:integralCurves}}
\end{figure}

The first Piola-Kirchhoff stress tensor derived from Eq.$\ $\eqref{eq:Gent}
is
\begin{equation}
\piola=\mu\left(1-\frac{\tr{\defgradT\defgrad}-3}{\jm}\right)^{-1}\defgrad+\left(\kappa-\frac{2\mu}{\jm}-\frac{2\mu}{3}\right)\det{\defgrad}\left(\det{\defgrad}-1\right)\defgradTi-\mu\defgradTi,\label{eq:gentstress}
\end{equation}
and the corresponding components required for calculating Eq.$\ $\eqref{eq:defineCoeff}
are 
\begin{equation}
\begin{aligned}\piolacomp{11} & =\mu\left(\pressurestrain+1\right)\left(1-\frac{\pressurestrain^{2}+\shearstrain^{2}+2\pressurestrain}{\jm}\right)^{-1}+\left(\kappa-\frac{2\mu}{\jm}-\frac{2\mu}{3}\right)\pressurestrain-\frac{\mu}{\pressurestrain+1},\\
\piolacomp{21} & =\mu\shearstrain\left(1-\frac{\pressurestrain^{2}+\shearstrain^{2}+2\pressurestrain}{\jm}\right)^{-1}.
\end{aligned}
\label{eq:pexplicit-1}
\end{equation}
The resultant characteristic velocities are 
\begin{equation}
c_{\pm}^{2}=\frac{\vc_{3}}{2\refden}+\frac{\mu}{2\refden\vc_{1}^{2}}+\frac{\left(\vc_{1}^{2}+\shearstrain^{2}+\jm\right)\mu\jm}{2\refden\vc_{2}^{2}}\pm\frac{\sqrt{48\vc_{1}^{6}\mu^{2}\shearstrain^{2}J_{m}^{2}+\left(\vc_{1}^{2}\vc_{2}^{2}\vc_{3}+2\vc_{1}^{4}\mu J_{m}-2\vc_{1}^{2}\shearstrain^{2}\mu\jm+\vc_{2}^{2}\mu\right){}^{2}}}{2\sqrt{3}\refden\vc_{2}^{2}\vc_{1}^{2}},\label{eq:cmgent-1}
\end{equation}
where 
\begin{equation}
\vc_{1}=\pressurestrain+1,\ \vc_{2}=\pressurestrain^{2}+2\pressurestrain+\shearstrain^{2}-\jm,\ \vc_{3}=\kappa-\frac{2\mu}{3}-\frac{2\mu}{\jm}.\label{eq:cg comp}
\end{equation}
In the calculations to follow, we use representative moduli of rubber
(see \citealp{Marckmann2006yu,Getz2017gsm}, and the references therein),
namely,
\begin{equation}
\kappa=1\,\mathrm{MPa},\ \mu=200\,\mathrm{kPa},\ \jm=10.\label{eq:paremeters}
\end{equation}
Numerical solutions of Eqs.$\ $\eqref{integralCurves}-\eqref{eq:comp0}
 are shown in Fig.$\ $\ref{fig:integralCurves}. We clarify that
these solutions are not associated with a concrete problem, and are
merely solutions of nonlinear differential equations with compatibility
conditions. The resultant curves can be identified with the evolution
of $\shearstrain$ and $\pressurestrain$ during the propagation of
smooth waves which spread in a uniformly pre-strained homogeneous
Gent medium, when subjected to impact \citep{ZIV2019mom}. Here, we
present the shear strain $\shearstrain$ as function of the axial
strain $\pressurestrain$, associated with the quasi-pressure (panel
a) and quasi-shear (panel b) waves. Each curve is obtained using different
values for the compatibility condition \eqref{eq:comp0}, which can
be identified by the intersection of a certain curve with the vertical
(panel a) and horizontal (panel b) axis. Accordingly, each curve in
panel (a) corresponds to one of the conditions 
\begin{equation}
\shearstrain\left(\pressurestrain=0\right)=1.7,1.8,...,2.5,\label{eq:specificcomp}
\end{equation}
where we note that the curves associated with the values $2,2.2$
and $2.4$ will we used in the sequel. Similarly, each curve in panel
(b) corresponds to one of the conditions 
\begin{equation}
\shearstrain\left(\pressurestrain=-0.4,-0.2,...,0.6\right)=0.\label{eq:specificcomp-1}
\end{equation}
We observe that for quasi-pressure waves $\shearstrain$ is a monotonically
increasing function of $\pressurestrain$, while for quasi-shear waves
$\shearstrain$ is a monotonically decreasing function of $\pressurestrain$.
\emph{It is important to note that the solutions to Eq.$\ $\eqref{integralCurves}
are independent of the initial mass density $\refden$}. This property
will be useful when solving the main problem, where the initial mass
density of the medium varies in space. \floatsetup[figure]{style=plain,subcapbesideposition=top}
\\

\emph{Main problem: laminated media}.---Having at hand the solutions
of the differential equations \eqref{etea} which are associated with
a homogeneous medium, we return to the main problem of the laminated
media. Specifically, consider a laminate whose layers are governed
by Eqs.$\ $\eqref{eq:Gent}-\eqref{eq:paremeters}, have an equal
thickness of $1\,$cm, and the initial mass density of phases $a$
and $b$ is $500\,\mathrm{kg/m^{3}}$ and $4000\,\mathrm{kg/m^{3}}$,
respectively. We specifically seek solutions to the response of the
laminate when subjected to non-uniform initial strains. The initial
strains we consider are in the form of a sum of a uniform part and
a localized part about $X_{1}=0$, namely,

\begin{equation}
\epsilon_{i}\left(X_{1},t=0\right)=\begin{cases}
\initial{\epsilon_{i}}+\amp{\epsilon_{i}}\cos\frac{\pi X_{1}}{w}, & -\frac{w}{2}<X_{1}<\frac{w}{2},\\
\initial{\epsilon_{i}}, & \mathrm{otherwise},
\end{cases}\label{eq:initialState}
\end{equation}
where $i=A,\,T$ denotes the axial and shear strains, respectively,
and superscripts $\left(U\right)$ and $\left(L\right)$ correspond
to the uniform and localized parts of the pre-strain, respectively.\floatsetup[figure]{style=plain,subcapbesideposition=top}
\begin{figure}[!t]
\sidesubfloat[]{\includegraphics[width=0.45\linewidth]{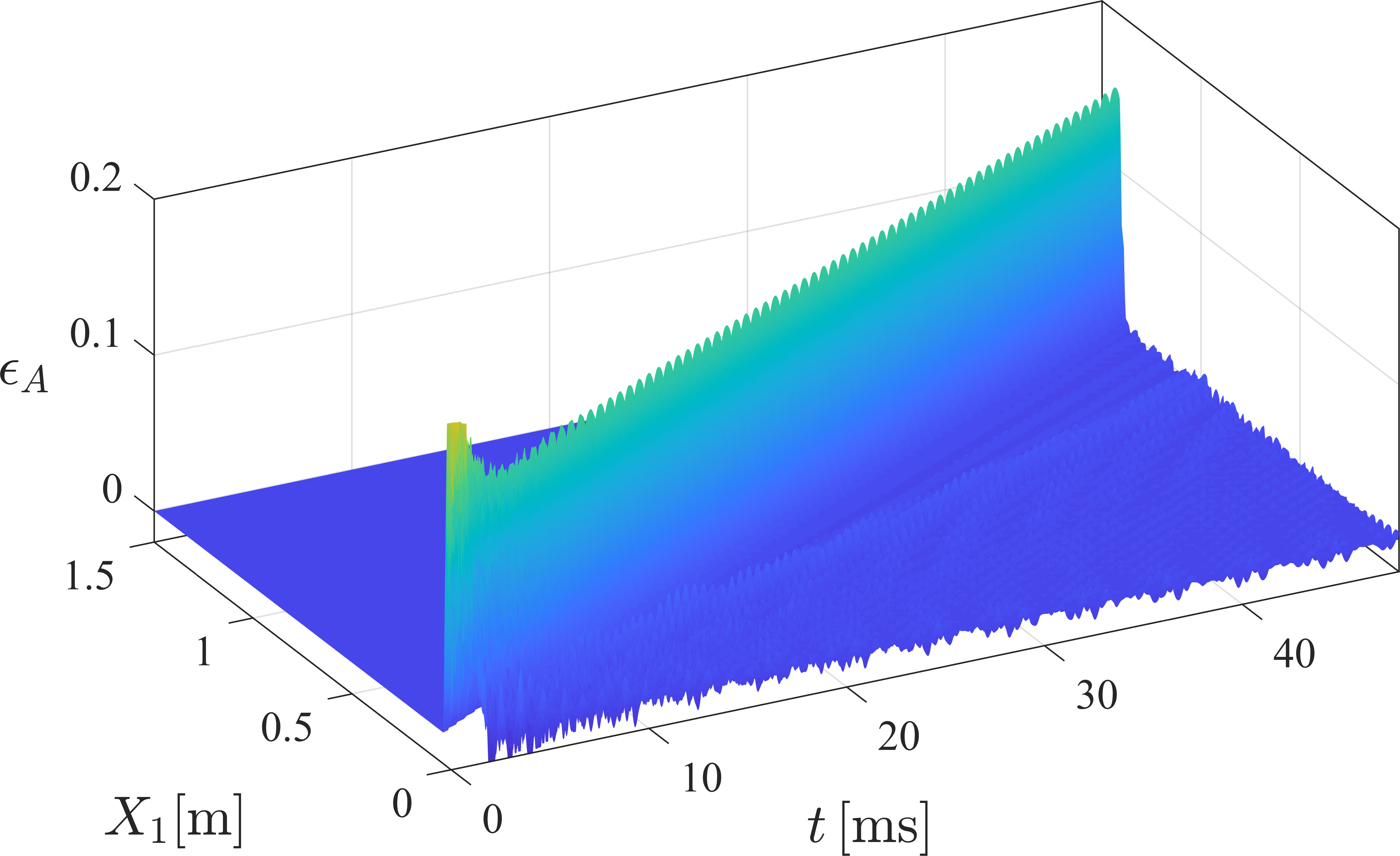}$\ \ \ $\includegraphics[width=0.45\linewidth]{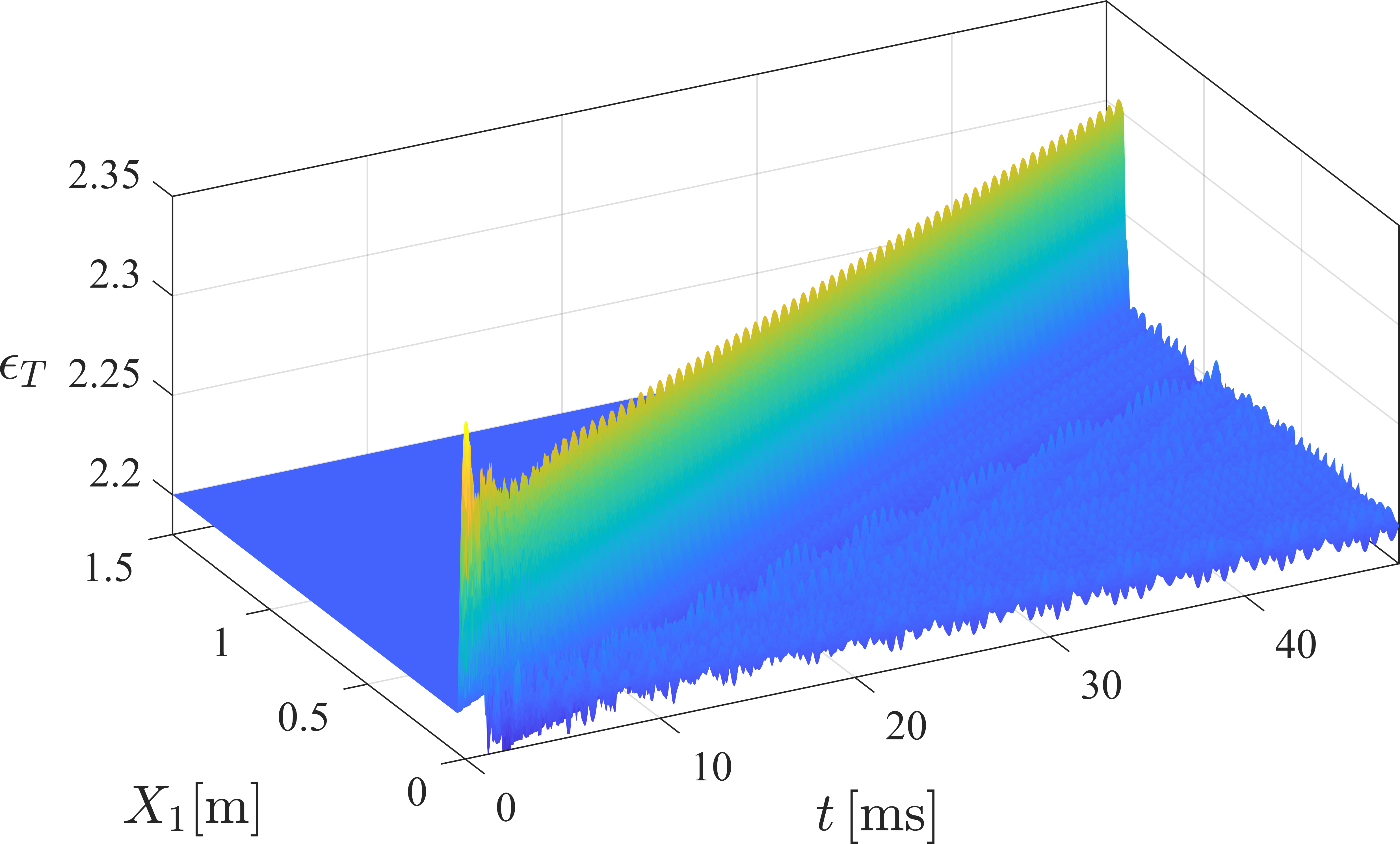}}  \\
\medskip{}\sidesubfloat[]{\includegraphics[width=0.45\textwidth]{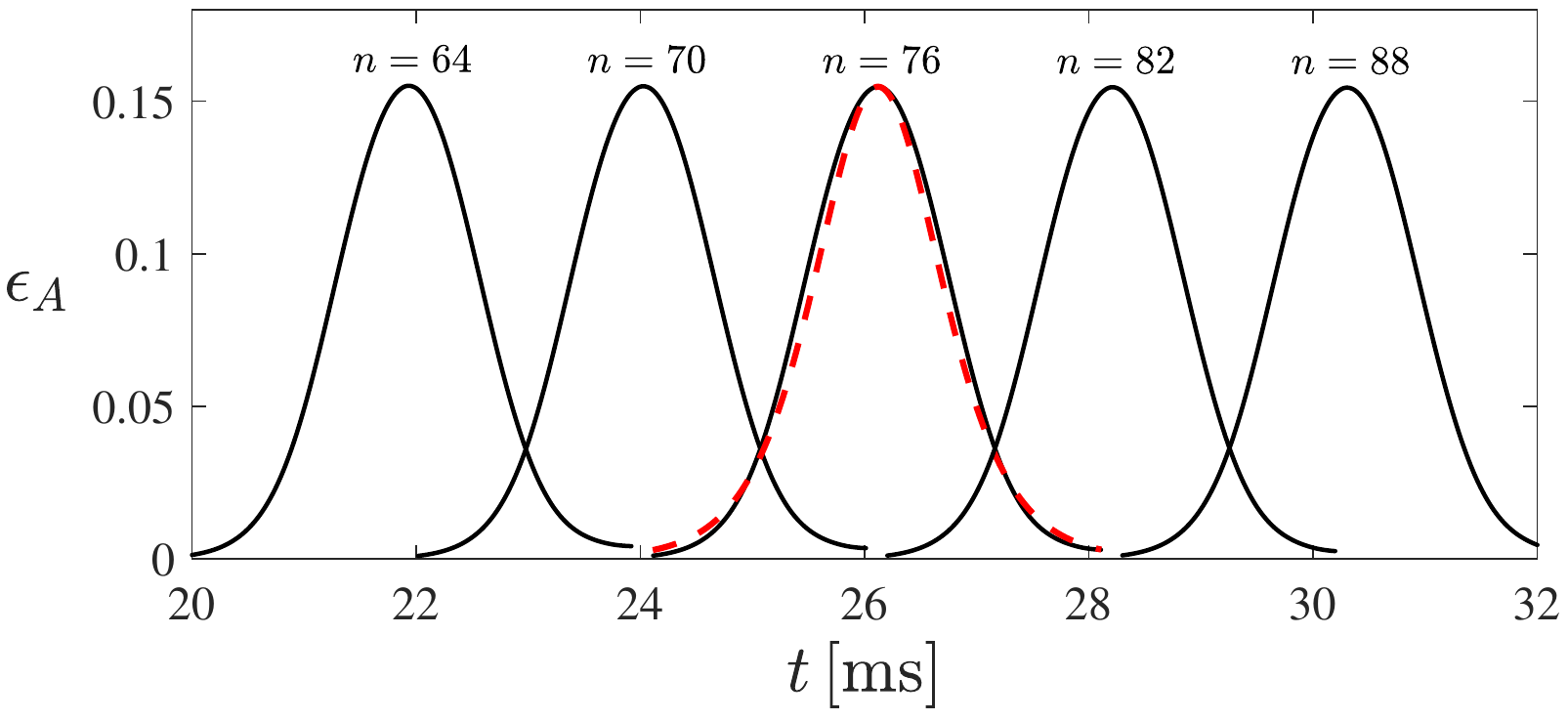}$\ \ \ $\includegraphics[width=0.45\textwidth]{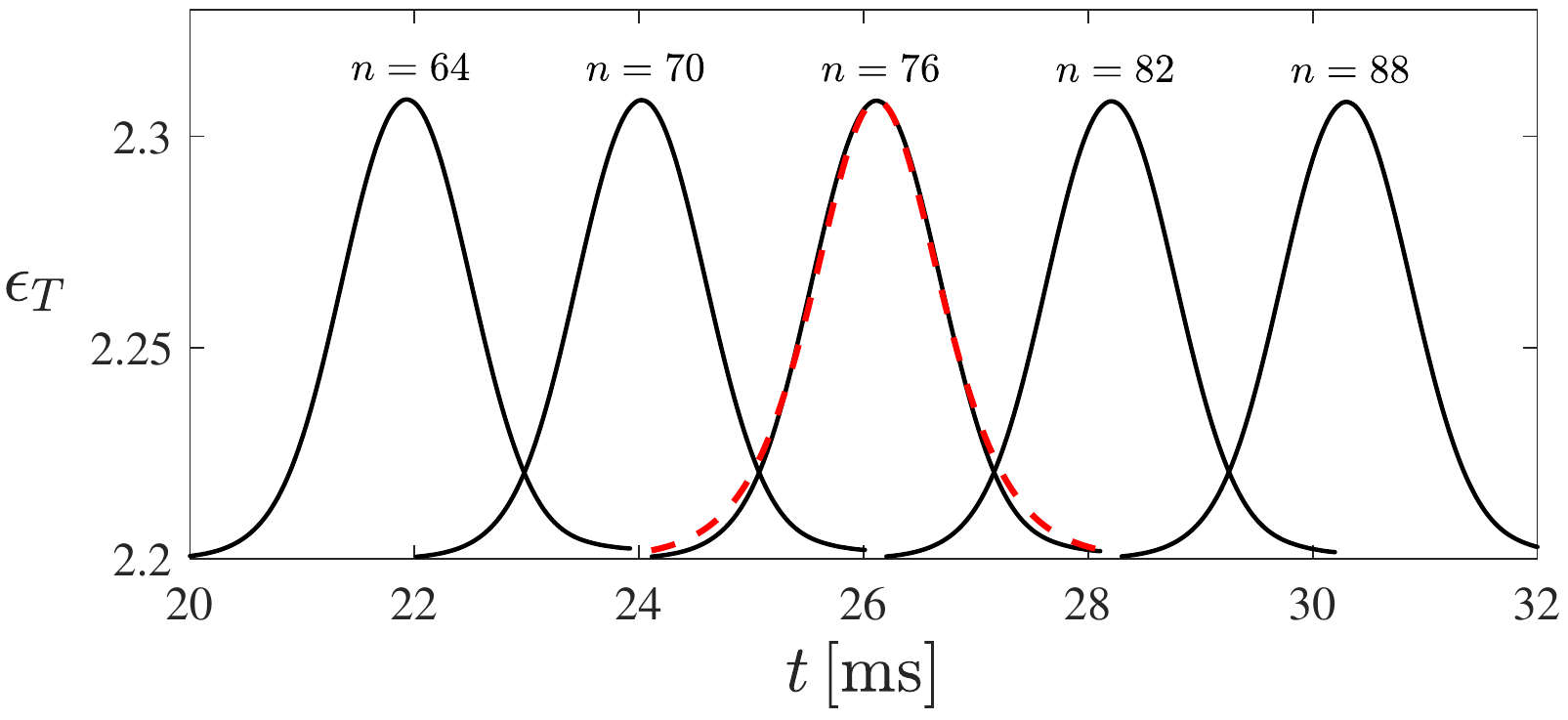}}  

\caption{(a) Axial (left panel) and shear (right panel) strains of a quasi-pressure
solitary wave as functions of space and time, when the laminate is
pre-strained according to Eq.$\ $\eqref{eq:quasi-pressure-strains}.
(b) Corresponding axial (left panel) and shear (right panel) strains
at the center of selected layers made of phase $a$, as functions
of time. Arbitrarily, we chose each $6^{\mathrm{th}}$ layer from
the range $64\protect\leq n\protect\leq88$; each curve corresponds
to a different layer whose number appears on top of the curve. The
red dashed curve depicts sech$^{2}$ functions given in Eq.$\ $\eqref{eq:sech-1}.\label{fig:pressureSoliton}}
\end{figure}
As mentioned in Sec.$\ $\ref{sec:Governing-equations}, to solve
the equations of motion we apply the finite-volume method of \citet{ziv2019b},
which was developed for analyzing motions comprising two coupled displacements
in soft laminates.

We begin by showing in Fig.$\ $\ref{fig:pressureSoliton}(a) the
distribution of the axial (left panel) and shear (right panel) strains
as functions of space and time, when the laminate is subjected to
a pre-strain with the parameters

\begin{equation}
\initial{\shearstrain}=2.2,\ \initial{\pressurestrain}=0,\ \amp{\shearstrain}=0.15,\ \amp{\pressurestrain}=0.3,\ w=0.08\,\mathrm{m}.\label{eq:quasi-pressure-strains}
\end{equation}
We observe that a rightward propagating wave is generated. A movie
of this propagation is also provided in the supplementary material
online under the name S1. Remarkably, while the wave profile is different
in phases $a$ and $b$, as it should, the respective profile in each
phase is maintained in passing from one unit cell to another. Furthermore,
by calculating the time of flight that takes the maximum of the strains
to pass from a certain phase in one periodic cell to the same phase
in adjacent cell, we find that its velocity is constant and equal
for both phases; this wave is therefore a solitary wave. More specifically,
since the solitary wave exhibits coupling between its axial and transverse
components and polarizations, it is a \emph{vector }solitary wave.
To the best of our knowledge, these vector solitary waves were first
observed by \citet{ziv2019b} using numerical experiments. Here, we
carry out a comprehensive characterization of the possible vector
solitary waves in the medium.

We first observe that the wave couples axial and shear strains of
an identical sign, similarly to quasi-pressure waves in a homogeneous
compressible Gent medium (Fig.$\ $\ref{fig:integralCurves}a). Hence,
we associate the solitary wave with a quasi-pressure mode, and term
it a quasi-pressure solitary wave. To facilitate the identification
of these features, we show in Fig.$\ $\ref{fig:pressureSoliton}(b)
the corresponding axial (left panel) and shear (right panel) strains
at the center of selected layers made of phase $a$, as functions
of time. Arbitrarily, we chose each $6^{\mathrm{th}}$ layer from
the range $64\leq n\leq88$; each curve corresponds to a different
layer whose number appears on top of the curve. Evidently, the shape
is maintained, and the equal spacing between the curves confirms that
the velocity is constant too. Furthermore, we observe that the wave
profile is reminiscent of the sech$^{2}$ profile, which, as mentioned,
is the solitary wave profile associated with the well-known KdV equation
\citep{korteweg1895xli}. This wave profile was also obtained for
one-dimensional solitary waves in a laminate, using an asymptotic
homogenization method \citep{andrianov2014numerical}. A comparison
between the wave profile at $n=76$ and the sech$^{2}$ functions
\begin{equation}
\pressurestrain=0.154\mathrm{sech}^{2}\left(1.33t\right),\ \shearstrain=2.2+0.108\mathrm{sech^{2}}\left(1.33t\right)\label{eq:sech-1}
\end{equation}
(plotted in red dash), demonstrates the agreement between them.

\floatsetup[figure]{style=plain,subcapbesideposition=top}
\begin{figure}[!t]
\sidesubfloat[]{\includegraphics[width=0.45\linewidth]{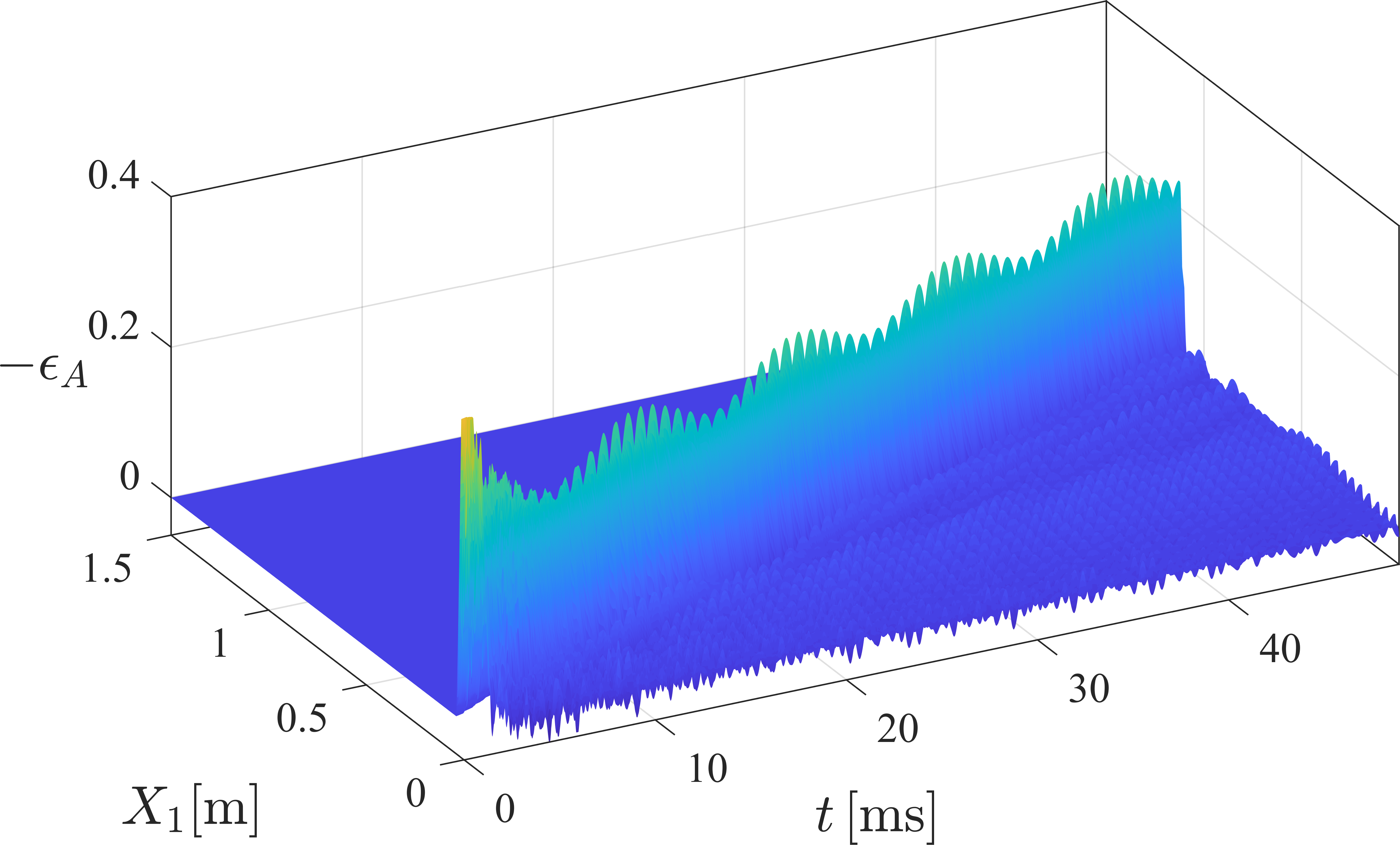}$\ \ \ $\includegraphics[width=0.45\linewidth]{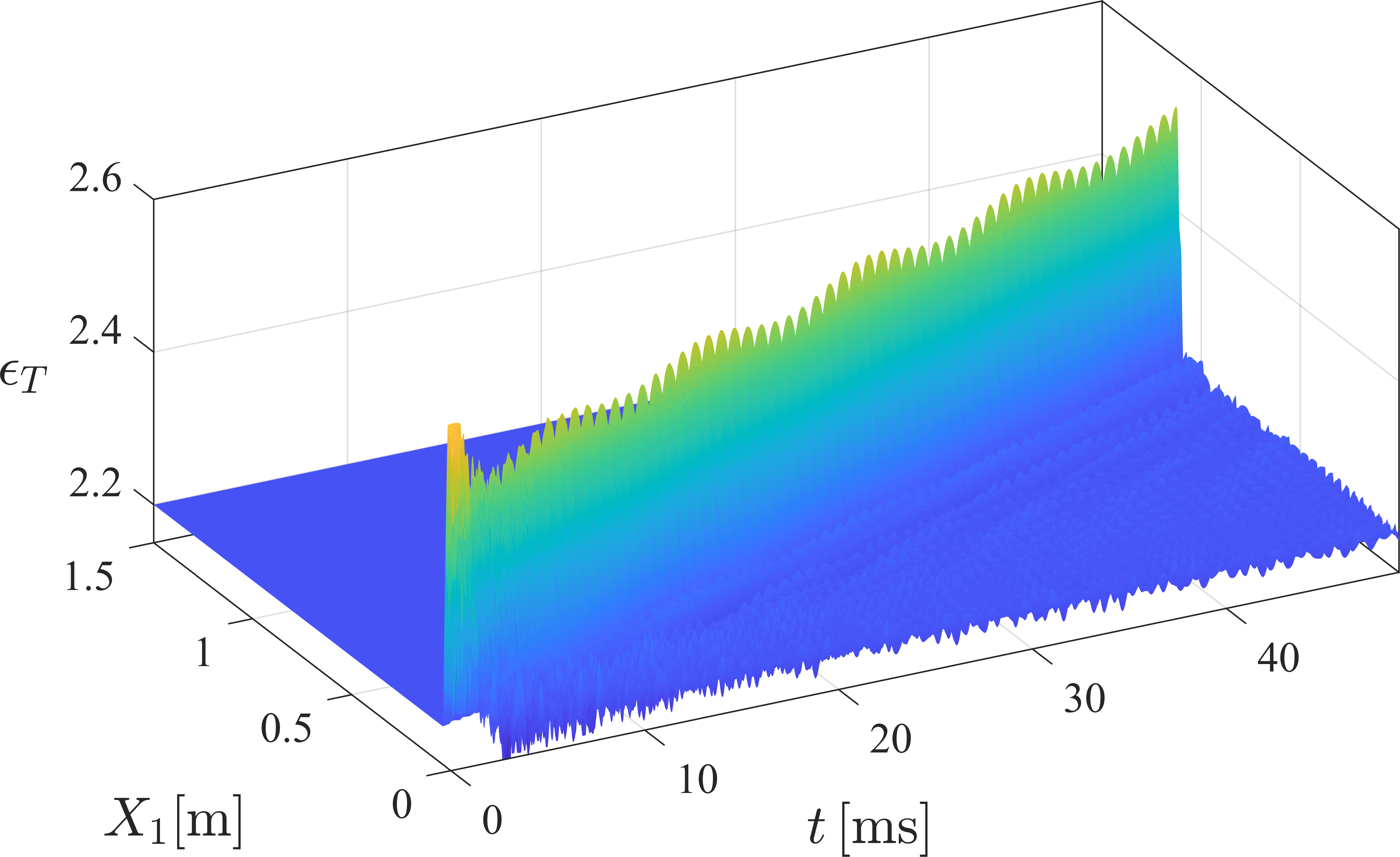}}  \\
\medskip{}\sidesubfloat[]{\includegraphics[width=0.45\textwidth]{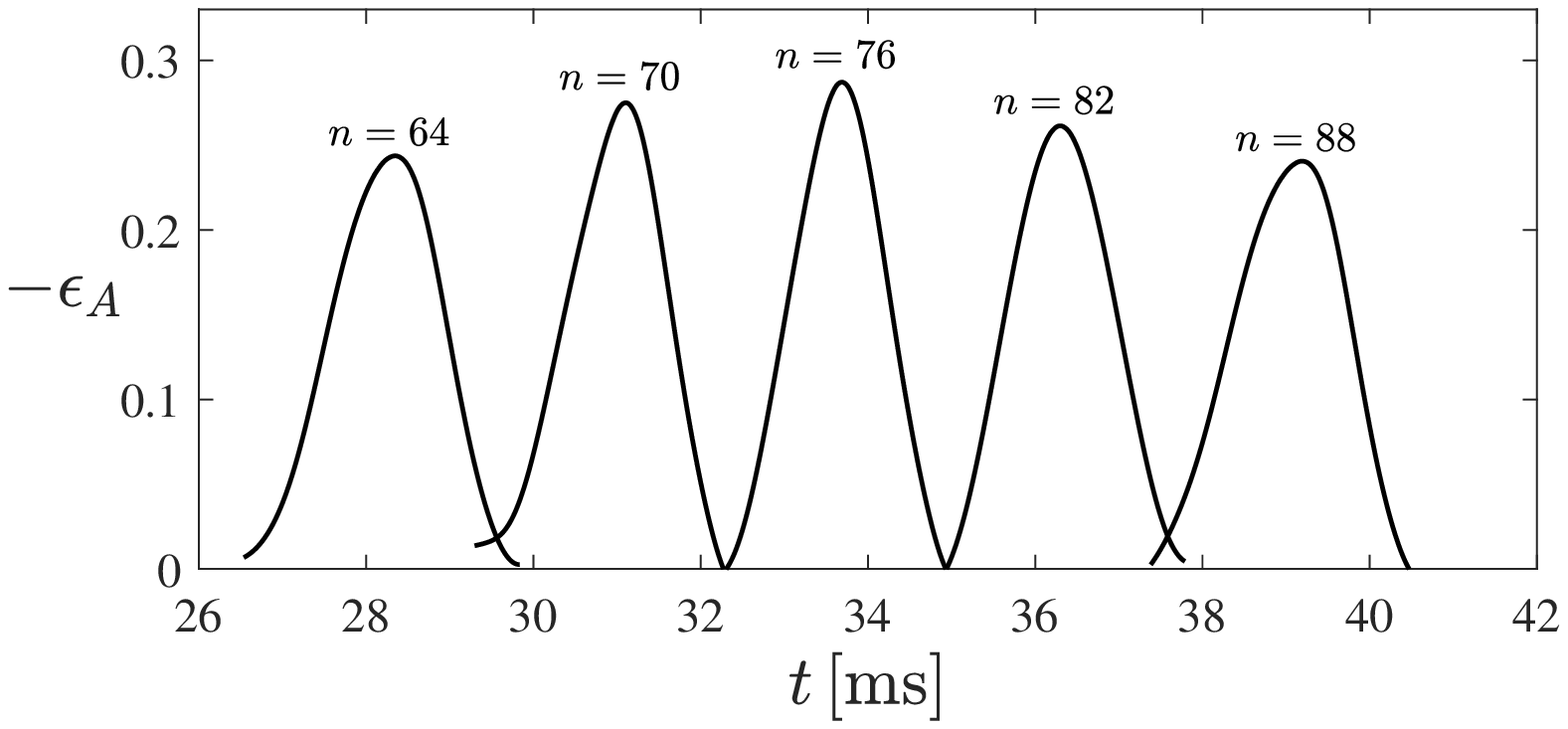}$\ \ \ $\includegraphics[width=0.45\textwidth]{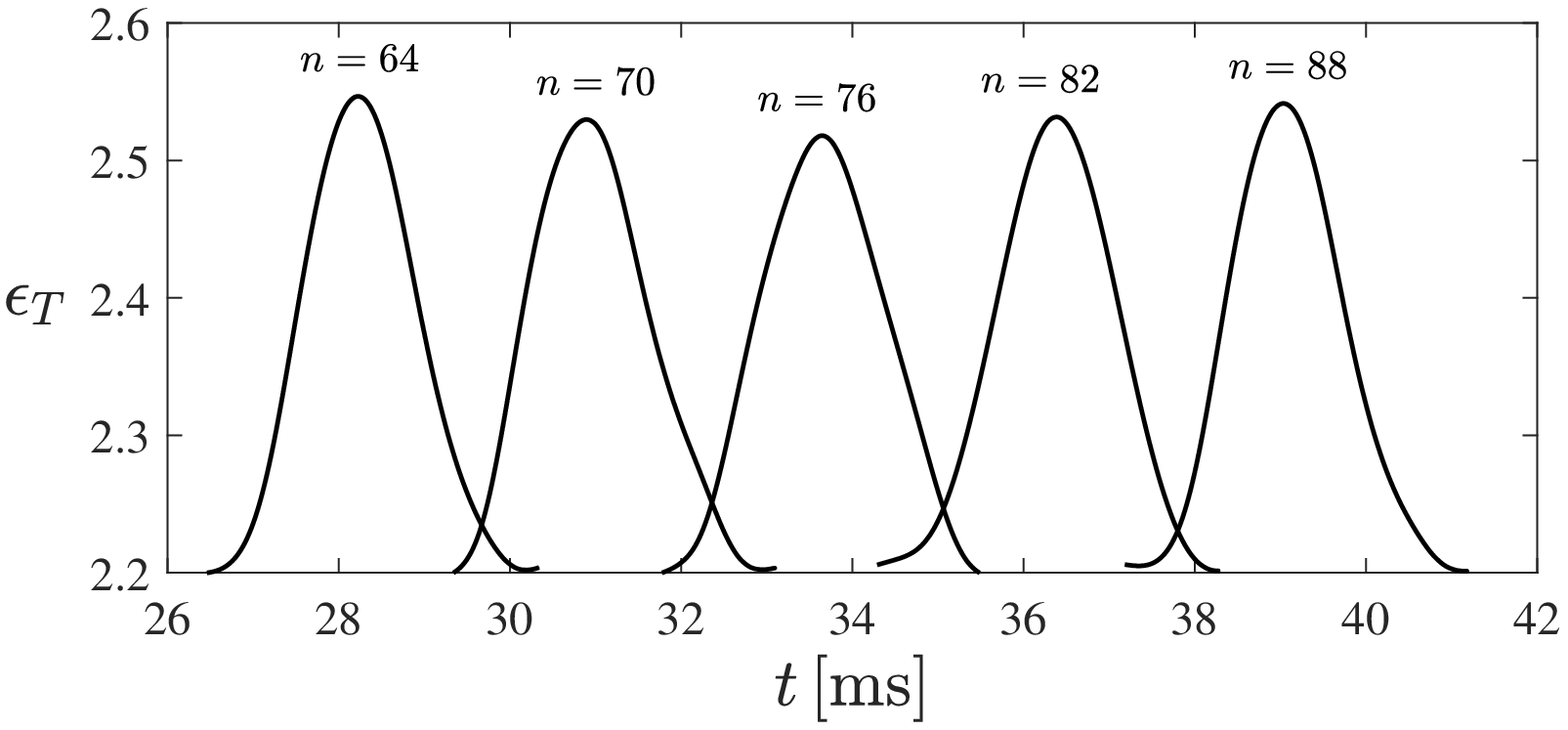}}  

\caption{(a) Axial (left panel) and shear (right panel) strains of a quasi-shear
solitary wave as functions of space and time, when the laminate is
pre-strained according to Eq.$\ $\eqref{eq:quasi-shear-strains}.
(b) Corresponding axial (left panel) and shear (right panel) strains
at the center of selected layers made of phase $a$, as functions
of time. Arbitrarily, we chose each $6^{\mathrm{th}}$ layer from
the range $64\protect\leq n\protect\leq88$; each curve corresponds
to a different layer whose number appears on top of the curve.\label{fig:SolitonShear}}
\end{figure}
Fig.$\ $\ref{fig:SolitonShear}(a) shows the distribution of the
axial (left panel) and shear (right panel) strains as functions of
space and time, when the laminate is pre-strained according to the
parameters

\begin{equation}
\initial{\shearstrain}=2.2,\ \initial{\pressurestrain}=0,\ \amp{\shearstrain}=0.5,\ \amp{\pressurestrain}=-0.5,\ w=0.08\,\mathrm{m}.\label{eq:quasi-shear-strains}
\end{equation}
We observe that a rightward propagating wave is generated, as also
depicted in movie S2 in the supplementary material, with the following
features. First, by examining the wave profile, we find that in contrast
with quasi-pressure waves, here (\emph{i}) the time dependency of
the waves across each layer is not a sech$^{2}$ function\textemdash nor
any other symmetric function; (\emph{ii}) the wave profile changes
not only between the different phases, but also between adjacent unit
cells. The latter change in the profile occurs in a periodic manner
without dispersing in time.

Second, by calculating the time of flight that takes the maximum of
the strains to pass from a certain phase in one periodic cell to the
same phase in adjacent cell, we find that the wave velocity oscillates
about a constant value. These persistent oscillations of the amplitude
and velocity occur through a periodic transfer of energy between the
axial and transverse strains of the wave, which we consider as internal
mode of the solitary wave. Similar long-lived oscillations were reported
before in other fields \citep{Campbell1983ty,Mantsyzov1995pra,kivshar1998internal,Szankowski2010prl},
and recently were also reported in an acoustic system using the continuum
limit of a discrete model based on mechanical\textendash electrical
analogies \citep{zhang2017bright}. Our results are the first report
of oscillating vector solitary waves that are generated from the equations
of continuum elastodynamics. 

Lastly, the axial and shear strains associated with the wave have
an opposite sign, similarly to quasi-shear waves in a homogeneous
compressible Gent medium (Fig.$\ $\ref{fig:integralCurves}b). On
account of these features, we term the wave\emph{ oscillating quasi-shear
vector solitary wave}\footnote{While the shape and velocity of the wave are not constant but rather
oscillate about a mean, we retain the term \emph{solitary}, in agreement
with the terminology used in other fields \citep{Mantsyzov1995pra,kivshar1998internal,Szankowski2010prl}.
This is due to the fact that the wave is stable, \emph{i.e.}, it does
not spread (or get squeezed) in the course of its propagation, in
spite of the fact that it is governed by nonlinear equations. This
also agrees with the interpretation of these oscillations as a result
of internal mode of a solitary wave. }\emph{.} 

In order to characterize the dependency of the wave velocity and the
relation between $\shearstrain$ and $\pressurestrain$ on the initial
strain, we perform a set of numerical experiments with different initial
states. Specifically, we set $\initial{\pressurestrain}=0$, consider
three values of $\initial{\shearstrain}$, namely, 2, 2.2 and 2.4
(denoted in the following figures by blue, red and green markers,
respectively), and for each value of $\initial{\shearstrain}$ we
carry out calculations considering 6 combinations of $\amp{\pressurestrain},\ \amp{\shearstrain}$
and $w$, chosen from the ranges

\begin{equation}
-0.8<\amp{\pressurestrain}<0.8,\ \ \ 0.1<\amp{\shearstrain}<0.7,\ \ \ 0.07\,\mathrm{m}<w<0.08\,\mathrm{m}.
\end{equation}
 These combinations were selected by trial and error to yield a single
vector solitary wave---either quasi-shear or quasi-pressure. By contrast,
the initial conditions used by \citet{ziv2019b} delivered trains
of solitary waves. The combinations for quasi-pressure solitary waves
are ordered such that when we increase $\amp{\shearstrain}$, we also
increase $\amp{\pressurestrain}$, where $\amp{\pressurestrain}$
is always positive. Conversely, the combinations for quasi-shear solitary
waves are ordered such that when we increase $\amp{\shearstrain}$,
we also decrease $\amp{\pressurestrain}$, where $\amp{\pressurestrain}$
is always negative. In both cases, the value of $w$ is decreased
when $\amp{\shearstrain}$ is increased.

\begin{figure}[!t]
\sidesubfloat[]{\includegraphics[scale=0.47]{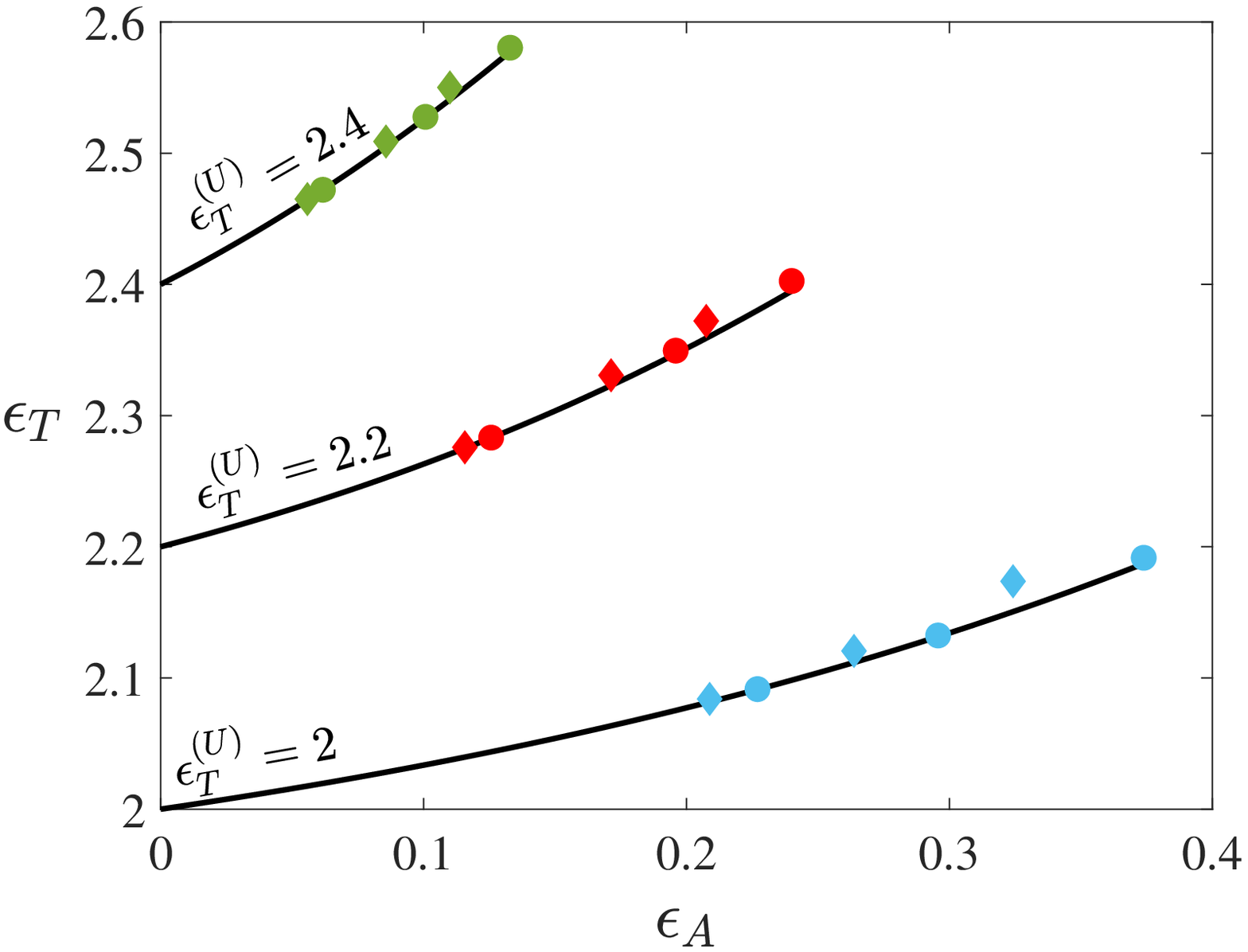}}  \sidesubfloat[]{\includegraphics[scale=0.47]{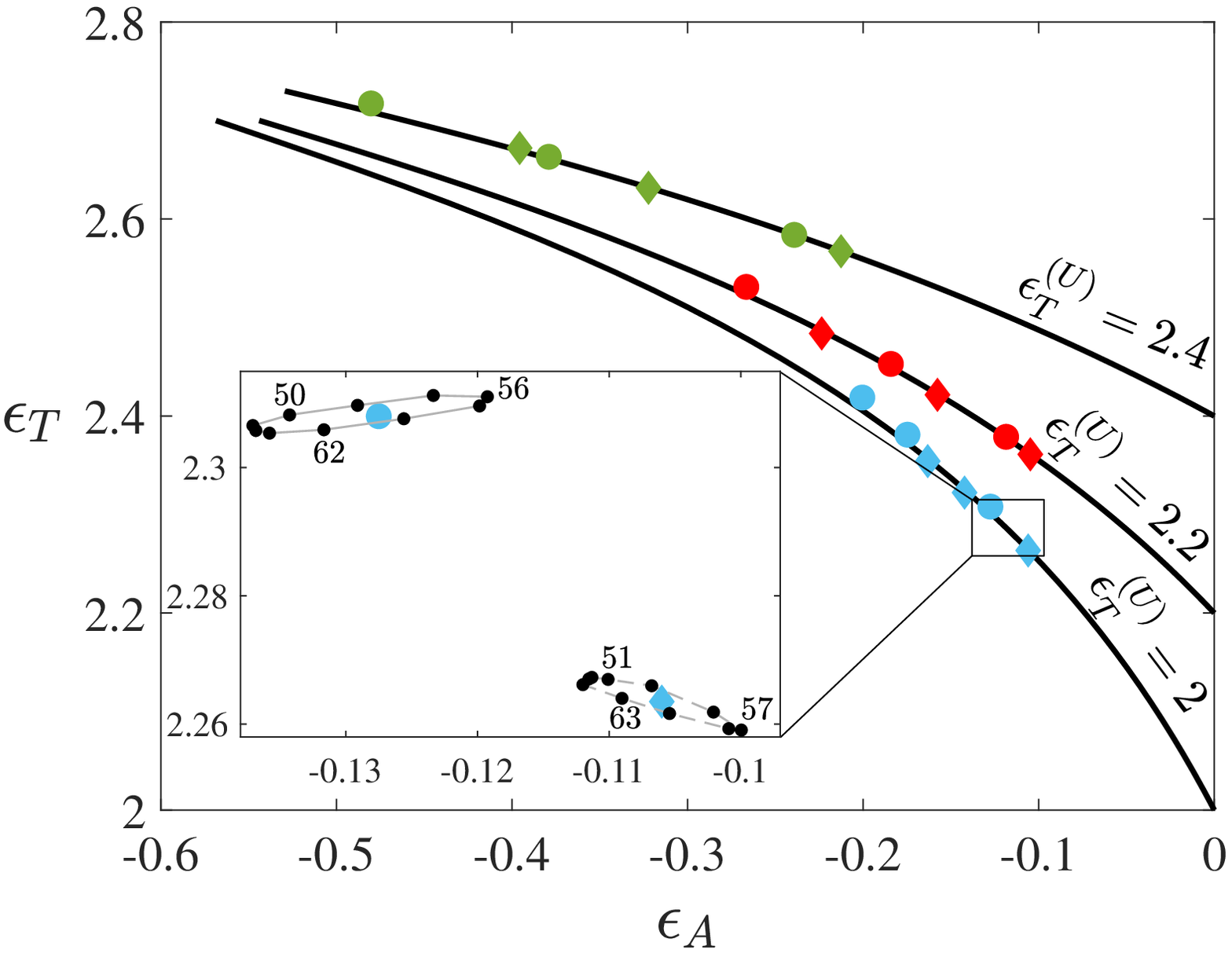}}  

\caption{$\protect\shearstrain-\protect\pressurestrain$ diagrams of vector
solitary waves in the soft laminate when subjected to $\protect\initial{\protect\pressurestrain}=0$
and $\protect\initial{\protect\shearstrain}=2,\,2.2$ and 2.4, depicted
by cyan, red and green markers, respectively. Each marker correspond
to a different initial localized strain. The black curves correspond
to the $\protect\shearstrain-\protect\pressurestrain$ relations in
a homogeneous compressible Gent material that is subjected to the
uniform part of the initial strain in the laminate. (a) Maximal values
of $\protect\shearstrain$ and $\protect\pressurestrain$ at the middle
of phase $a$ (discs) and $b$ (diamonds) over the time that the quasi-pressure
solitary wave traverses a unit cell. (b) Maximal values of $\protect\shearstrain$
and $\protect\pressurestrain$ over the time that the quasi-shear
solitary wave traverses a unit cell, averaged over the number of layers
within one period of the modulation. The values over which the average
was obtained are shown for two exemplary points in the inset; each
point corresponds to a different layer, where the number of three
arbitrary layers is displayed.\label{fig:Characterization}}
\end{figure}

Our results for quasi-pressure solitary waves are shown in Fig.$\ $\ref{fig:Characterization}(a),
in the following manner. From each simulation we extract the maximal
state of strain over the time that it takes the generated wave to
cross a unit cell. We specifically extract the maximal strain state
at the middle of phases $a$ and $b$, respectively. The corresponding
states are denoted in the $\left(\shearstrain,\pressurestrain\right)$
plane by disc (phase $a$) and diamond (phase $b$) markers. 

In addition to the results from the full-wave simulations, we also
present in Fig.$\ $\ref{fig:Characterization}(a) the solutions of
Eq.$\ $\eqref{eq:integralCurvesPressure} for different compatibility
conditions \eqref{eq:comp0}, as previously depicted in Fig.$\ $\ref{fig:integralCurves}.
While these solutions are not associated with a concrete problem,
they capture the way in which $\shearstrain$ and $\pressurestrain$
evolve during the propagation of smooth waves which spread in a uniformly
pre-strained homogeneous Gent half-space. We specifically present
using black curves the solutions to compatibility conditions \eqref{eq:comp0}
that agree with the uniform part of the initial strain in the laminate
($\amp{\pressurestrain}=\amp{\shearstrain}=0$), namely, 
\begin{equation}
\shearstrain\left(\pressurestrain=0\right)=\initial{\shearstrain},\quad\initial{\shearstrain}=2,2.2,2.4.\label{eq:initial conditions}
\end{equation}
 We observe that the results extracted from the full-wave simulations
of the laminate either coincide or fall very close to the curves that
are associated with the impact response of the homogeneous medium.
Specifically, we observe that the results for layers made of phase
$a$ coincide with these curves for all values of $\initial{\shearstrain}$.
We find it remarkable that these two seemingly unrelated problems
are related in this way. This agreement between the full-wave simulations
and the solution of Eq.$\ $\eqref{integralCurves} indicates that
the analysis of the homogeneous medium can be used to predict quasi-pressure
solitary waves in a soft laminate with a periodic modulation in its
mass density. This result relies on the fact that Eq.$\ $\eqref{integralCurves}
is independent of the mass density, and hence turns useful in the
solution of a laminate whose only modulation is in that property. 

\floatsetup[figure]{style=plain,subcapbesideposition=top}
Fig.$\ $\ref{fig:Characterization}(b) mirrors the presentation in
panel a for quasi-shear solitary waves. Since in this case the profile
changes between different cells, we present our results in the following
way. First, we have calculated the maximal values of $\shearstrain$
and $\pressurestrain$ over the time that the solitary wave traverses
a unit cell, as we previously did for the quasi-pressure waves. Since
now different layers exhibit different maximal values owing to the
internal mode, we here present the average value. Specifically, the
average of the values for phase $a$ (resp.$\ $$b$) is taken over
the number of layers within one period of the oscillation; the average
is denoted by the disc (resp.$\ $diamond) markers. As in panel (a),
to the results from the full-wave simulations we add in black curves
the solutions of Eq.$\ $\eqref{eq:integralCurvesShear}, when subjected
to the compatibility conditions \eqref{eq:initial conditions}. Here
again, the markers associated to the laminate either coincide or fall
very close to these curves.

To highlight the oscillatory nature of the wave amplitude, we present
in the inset of Fig.$\ $\ref{fig:Characterization}(b) the values
over which the average was obtained for two exemplary markers; each
point in the inset corresponds to a different layer that was used
in the calculation of the average, where the number of three arbitrary
layers is displayed. The slow spatial modulation in the wave profile
that appears in Fig.$\ $\ref{fig:SolitonShear} manifests itself
here as closed ellipses, showing how the amplitude oscillates periodically
about a mean. Interestingly, the orientations of the ellipses associated
with each phase are opposite one to another.

\begin{figure}[!t]
\sidesubfloat[]{\includegraphics[scale=0.45]{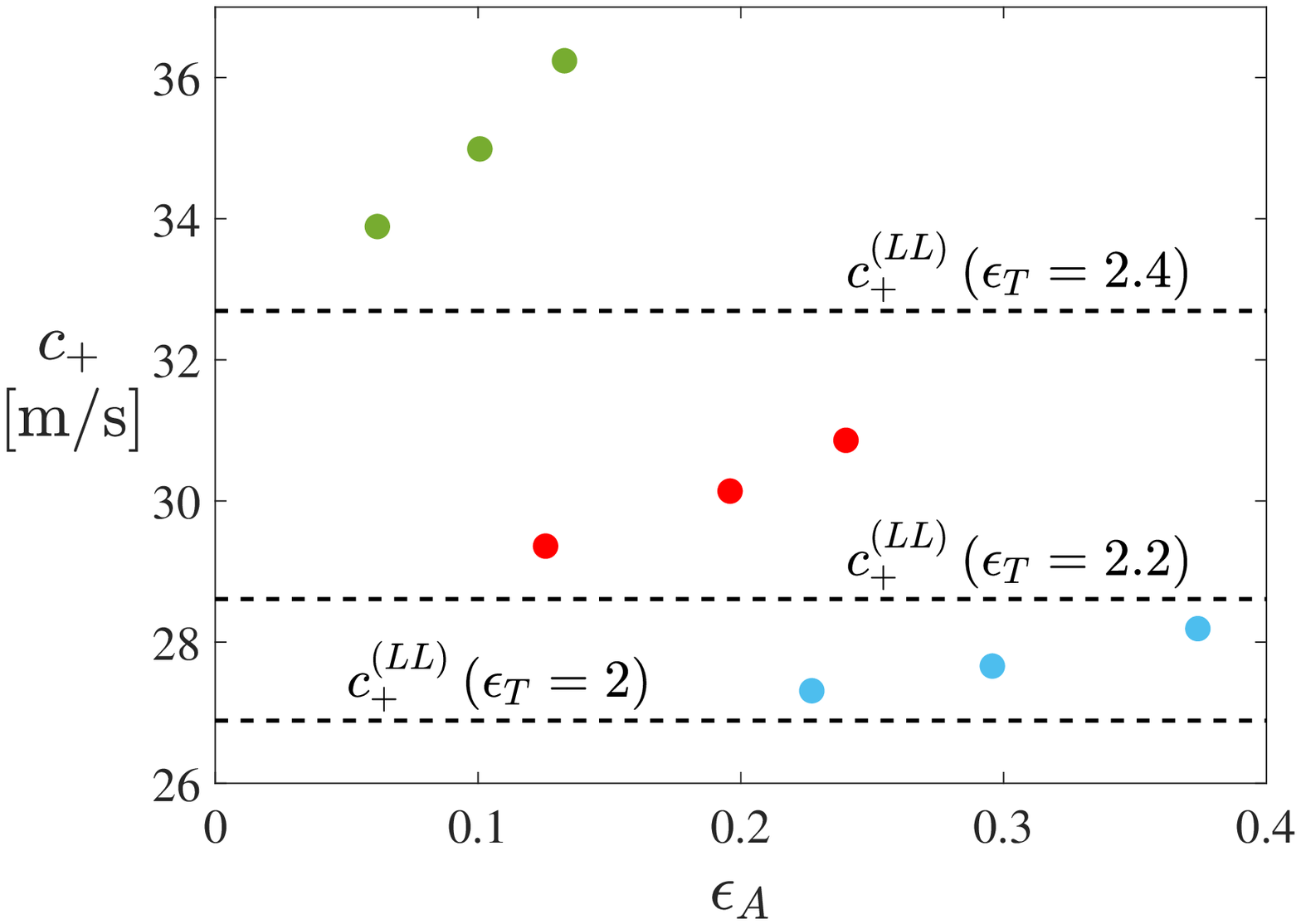}}  \sidesubfloat[]{\includegraphics[scale=0.45]{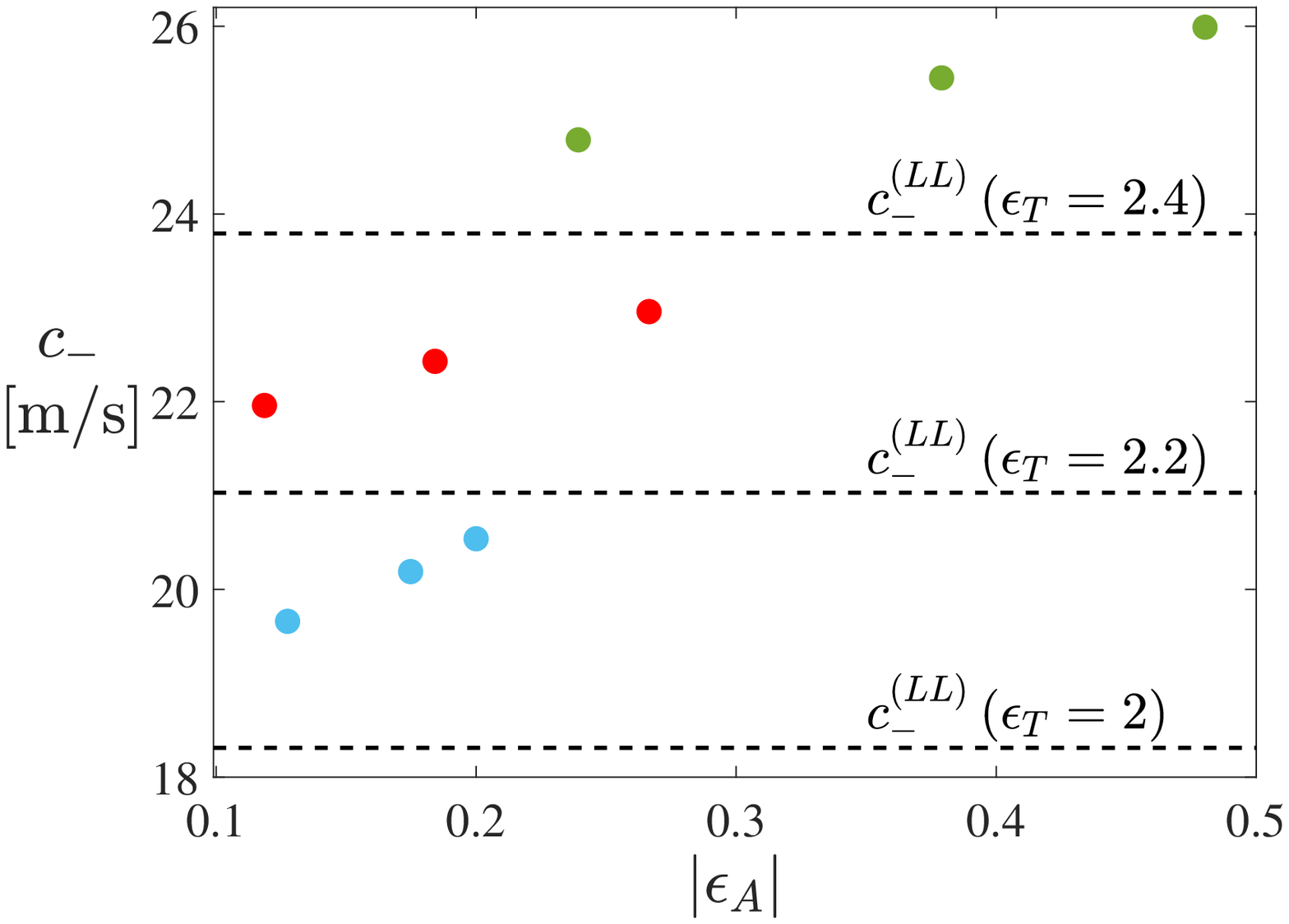}}  

\caption{(a) Velocity of quasi-pressure solitary waves as function of the maximal
value of $\protect\pressurestrain$ in phase $a$ over the time that
the solitary wave traverses a unit cell. (b) Velocity of quasi-shear
solitary waves as function of the maximal values of $\left|\protect\pressurestrain\right|$
in phase $a$ over the time that the solitary traverses a unit cell,
averaged over the number of layers within one period of the modulation.
The cyan, red and green discs correspond to solitary waves generated
with $\protect\initial{\protect\shearstrain}=2,\,2.2$ and $2.4$,
respectively. The dashed lines correspond to the values of $c_{\pm}^{\left(LL\right)}$
at each pre-strain.\label{fig:velocities}}
\end{figure}

The dependency of the velocity on the amplitude is shown in Fig.$\ $\ref{fig:velocities}(a)
by plotting the velocity of quasi-pressure solitary waves as function
of the maximal value of $\pressurestrain$ in phase $a$ over the
time that the solitary wave traverses a unit cell. Here again, the
cyan, red and green markers correspond to waves that were generated
with $\initial{\shearstrain}=2,2.2$ and $2.4$, respectively. We
observe that the highest and lowest clusters correspond to the largest
and smallest $\initial{\shearstrain}$, respectively. We furthermore
observe that within each cluster, markers at larger values of $\epsilon_{A}$
exhibit faster velocities. Accordingly, we conclude that the velocity
is a monotonically increasing function of $\shearstrain$ and $\pressurestrain$.
This is in agreement with the observation of \citet{ziv2019b} who
generated trains of solitary waves, and found that the taller waves
in the train propagate faster than shorter waves. This dependency
is the opposite of the dependency of the vector solitary waves observed
in the mechanical system conceived by \citet{deng2017}, as the two
problems are fundamentally different. Notably, our system is a continuum
solid which is constitutively and geometrically nonlinear, while the
system of \citet{deng2017} is captured by a model of linear springs
and rigid squares. Moreover, the physical properties of our continuum
are heterogeneous, as its initial mass density varies in space, while
in the system of \citet{deng2017} all the springs and squares share
the same properties. 

We show next that it is possible to bound from below the velocities
$c_{+}$ and $c_{-}$ using the velocities in the linear limit (Bloch-Floquet
waves) at the low-frequency, long-wavelength regime. Before we proceed,
we note that \citet{yong2003solitary} have numerically found that
the velocity of Bloch-Floquet compression waves in that regime is
a lower bound for the velocity of one-dimensional compression waves
in the nonlinear laminates they studied. \citet{andrianov2014numerical}
later showed this result analytically for a different nonlinear model,
using an asymptotic homogenization method \citep{andrianov2013dynamic}.
Here, we generalize their conclusion to the present vectorial problem,
where there are two coupled components of the displacement field and
two coupled wave polarizations, using the limiting velocities of the
quasi-shear and quasi-pressure waves. To this end, we follow \citet{santosa1991dispersive},
who rigorously showed that the velocity of the leading term in the
Bloch expansion is simply the root square of the harmonic average
of the stiffness over the weighted average of the mass density, and
further showed that this result agrees with multi-scale asymptotics
or homogenization. To carry out this calculation in the linear limit
of our problem, we evaluate the instantaneous stiffness of the phases
in the pre-strained configuration \citep{ogden97book,LUSTIG2018jmps},
and use the initial mass density. We denote the velocities that are
calculated using the resultant average for the quasi-pressure and
quasi-shear waves by $c_{+}^{\left(LL\right)}$ and $c_{-}^{\left(LL\right)}$,
respectively. These values at each pre-strain are given by the dashed
lines in Fig.$\ $\ref{fig:velocities}, confirming they are lower
bounds. This observation implies that quasi-pressure and quasi-shear
vector solitary waves are supersonic, thereby generalizing the results
of \citet{yong2003solitary} and \citet{andrianov2014numerical} in
the scalar case.

Fig.$\ $\ref{fig:velocities}(b) shows the velocity of quasi-shear
solitary waves as function of the maximal values of $\left|\pressurestrain\right|$
(recall that here $\pressurestrain<0$) in phase $a$ over the time
that the solitary wave traverses a unit cell, averaged over the number
of layers within one period of the modulation. Here again, the velocity
of quasi-shear solitary waves is a monotonically increasing function
of the strain amplitude, and is bounded from below by $c_{-}^{\left(LL\right)}$.

\begin{figure}[!t]
\sidesubfloat[]{\includegraphics[scale=0.35]{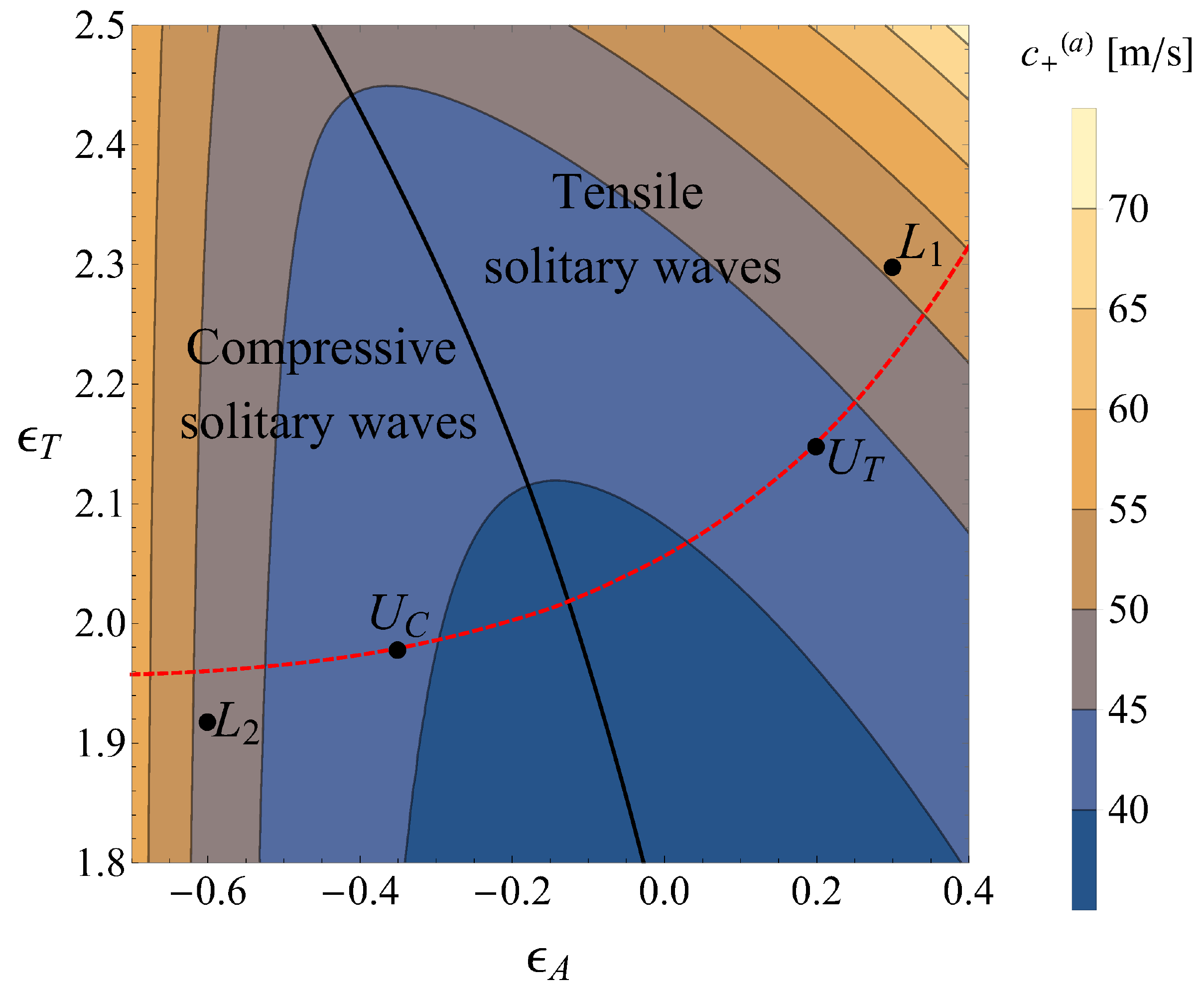}}  \sidesubfloat[]{\includegraphics[scale=0.35]{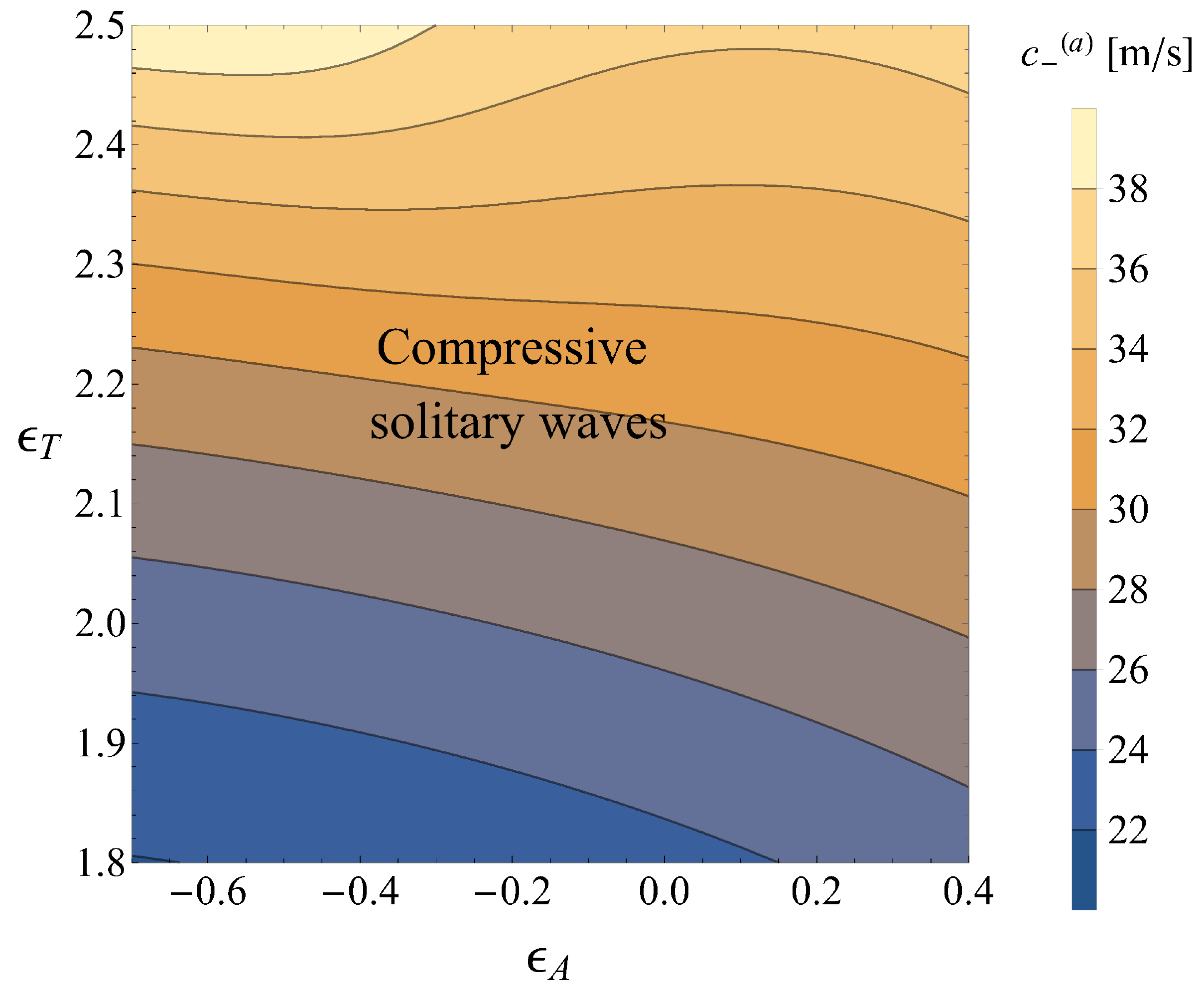}}  

\caption{Contour plots of (a) $\protect\phaseA{\protect\pressurewavespeed}$
and (b) $\protect\phaseA{\protect\shearwavespeed}$ as functions of
$\protect\pressurestrain$ and $\protect\shearstrain$. The solid
black curve in panel (a) separates $\left(\protect\initial{\protect\shearstrain},\,\protect\initial{\protect\pressurestrain}\right)$
pairs that support tensile solitary waves from pairs that support
compressive solitary waves. The markers denoted by $U_{C},\ U_{T},\ L_{1}$
and $L_{2}$ correspond to exemplary sets of two uniform strains and
two localized strains. respectively. The dashed red curve is the solution
of Eq.$\ $\eqref{eq:integralCurvesPressure} with either $U_{C}$
or $U_{T}$ as the initial condition.\label{fig:quasi-pressure-solitary}}
\end{figure}

\floatsetup[figure]{style=plain,subcapbesideposition=top}
We recall that for solitary waves to propagate there should be a balance
between dispersion and nonlinearity. In our case of a finitely deformed
laminate, this balance is achievable by pre-strain tuning: solitary
waves can propagate only at specific initial deformations for which
the material stiffens along the loading path \citep{hussein2018nonlinear}.
The quantification of this stiffening is given by the gradient of
the characteristic velocity along this path. Specifically in our case,
this condition is satisfied when each characteristic velocity, \emph{i.e.},
$\phaseA{c_{\pm}}$ and $\phaseB{c_{\pm}}$, is increasing monotonically
along the loading path given in Eqs.$\ $\eqref{eq:integralCurvesPressure}
and \eqref{eq:integralCurvesShear} for quasi-pressure and quasi-shear
solitary waves, respectively. Since in our exemplary laminate the
ratio between $\phaseA{c_{\pm}}$ and $\phaseB{c_{\pm}}$ is kept
constant, it is sufficient to characterize the dependency of $\phaseA{c_{\pm}}$
on the pre-strain. This dependency is characterized in Fig.$\ $\ref{fig:quasi-pressure-solitary},
where contour plots of $\phaseA{\pressurewavespeed}$ (panel a) and
$\phaseA{\shearwavespeed}$ (panel b) are shown as functions of $\left\{ \pressurestrain,\,\shearstrain\right\} $
pairs. We demonstrate how these maps are used to determine if a solitary
wave will propagate using a representative example, by considering
two different sets of uniform strains and two different sets of localized
strains, namely,\floatsetup[figure]{style=plain,subcapbesideposition=top}
\begin{equation}
\begin{aligned} & U_{C}=\left(\initial{\pressurestrain}=-0.35,\ \initial{\shearstrain}=1.98\right),\ \ \ U_{T}=\left(\initial{\pressurestrain}=0.2,\ \initial{\shearstrain}=2.15\right),\\
 & L_{1}=\left(\amp{\pressurestrain}=0.3,\ \amp{\shearstrain}=2.3\right),\ \ \ L_{2}=\left(\amp{\pressurestrain}=-0.6,\ \amp{\shearstrain}=1.92\right).
\end{aligned}
\end{equation}
The dashed red curve is the solution of Eq.$\ $\eqref{eq:integralCurvesPressure}
with either $U_{C}$ or $U_{T}$ as the initial condition. As mentioned
previously, the $\shearstrain-\pressurestrain$ relation of quasi-pressure
solitary waves generated with inhomogeneous pre-strain whose uniform
part is the same as $U_{C}$ (or $U_{T}$) approximately follows this
curve. Furthermore, the relative location of the localized part of
the strain with respect to the uniform strain determines the direction
of the strain evolution along the red curve, such that $L_{1}$ and
$L_{2}$ correspond to an increase and a decrease in $\pressurestrain$,
respectively. When the initial strain in the laminate is the sum of
the uniform strain $U_{T}$ and the localized strain $L_{1}$, we
have that $\phaseA{\pressurewavespeed}$ is monotonically increasing
along the dashed red curve, whose general direction is towards $L_{1}$.
Therefore, this initial state supports a quasi-pressure solitary wave.
When the initial strain is the sum of $U_{T}$ and $L_{2}$, we have
that $\phaseA{\pressurewavespeed}$ is monotonically decreasing along
the dashed red curve, whose general direction is towards $L_{2}$.
Thus, this initial state does not support solitary waves and the generated
wave will disperse. Similarly, when the uniform part of the initial
strain is given by $U_{C}$ and the localized part is $L_{1}$, solitary
waves cannot propagate, whereas when the localized part is $L_{2}$
such waves do propagate.  We emphasize that without the observation
that the $\shearstrain-\pressurestrain$ relations of the laminate
approximately follow their counterparts in the homogeneous case, it
would have not been possible to estimate \emph{a priori} what should
be the localized strains that generate a certain solitary wave. 

These contour maps are also useful in predicting how the axial strain
changes during the propagation of the solitary waves, \emph{i.e.,
}will it decrease increase. If the former occurs, we term the wave
a compressive wave, owing to its tendency to decrease the axial strain
in the laminate, notwithstanding the fact that the sign of axial strain
in the laminate may be positive. Similarly, if the latter occurs,
we term the wave a \emph{tensile} solitary wave, notwithstanding the
fact that the sign of axial strain may be negative. For compressive
solitary waves to form, it is required that along the loading path
of the solitary wave, the value of $\pressurestrain$ is decreasing
and the phase velocity is monotonically increasing. Specifically for
quasi-pressure waves, the domain for which this is possible is identified
with an initial uniform strain that belongs to the domain left to
the black curve in Fig.$\ $\ref{fig:quasi-pressure-solitary}(a).
Conversely, the formation of tensile solitary waves requires that
along the loading path of the solitary wave, the value of $\pressurestrain$
is increasing and the phase velocity is monotonically increasing.
The domain of initial uniform strains that can satisfy this criterion
is to the right of the curve. The condition for tensile waves cannot
be satisfied for quasi-shear solitary waves where only compressive
solitary waves are supported, independently of the initial strain.

Interestingly, the prescribed strains supporting the formation of
solitary waves in laminates and shock waves in homogeneous media are
similar, as they both require that the wave velocity will increase
monotonically along the loading path. The latter problem of shocks
in compressible Gent materials was studied by \citet{ZIV2019mom}
when considering coupled shear-pressure shocks, and later by \citet{CHOCKALINGAM2020103746}
who restricted attention to shear shocks in the specialized incompressible
case. The configuration considered by \citet{ZIV2019mom} consists
of a pre-strained semi-infinite Gent material with the same model
parameters as our laminate, namely, Eq.$\ $\eqref{eq:paremeters},
subjected to an impact at the boundary. The analogy between the problems
is manifested by the similarity of the prescribed strains supporting
solitary waves and shock waves. Specifically, if the magnitude of
the localized part of the initial strain is substituted as the magnitude
of the impact at the boundary of the homogeneous half-space, then
Fig.$\ $\ref{fig:quasi-pressure-solitary} also predicts which impacts
yield tensile shock waves and which impacts yield compressive shock
waves. In other words, impacts that give rise to monotonically increasing
phase velocities and in turn shocks in a homogeneous Gent material,
also give rise to monotonically increasing phase velocities and in
turn may form solitary waves in laminates made of the same Gent material
whose mass density is modulated. This connection sheds light on the
balance between the nonlinearity in the system and its dispersion:
the constitutive nonlinearity causes the pulse to steepen, while the
scattering at the interfaces disperses it and generates solitary waves
\citep{hussein2018nonlinear}, otherwise the steepening would have
result in shock formation \citep{yong2003solitary}.

\section{Summary\label{sec:Summary}}

\citet{ziv2019b} have developed a finite-volume method to solve the
equations governing finite-amplitude smooth waves with two coupled
components in nonlinear compressible laminates. The application of
this method to simulate the response of pre-strained compressible
Gent two-phase laminates has revealed the generation of vector solitary
waves, whose axial and transverse polarizations are coupled. Here,
we have used the method to conduct a large  set of numerical experiments
with different initial conditions, in order to characterize these
waves. 

We have classified the possible vector solitary waves according to
two types of profiles. The first class\textemdash which was observed
by \citet{ziv2019b}\textemdash has a profile that varies between
the two phases, but remains fixed as the wave passes through the different
periodic cells. We have termed this type as quasi-pressure, since
it reduces to the standard pressure waves in the limit of small strains
and vanishing heterogeneity, and showed that its components follow
the profile of the sech$^{2}$ function. 

The second class we have identified was not generated by \citet{ziv2019b},
and is more interesting; we termed it quasi-shear with similar reasoning.
The profile and velocity of this class varies not only between different
phases, but also between different cells in a periodic manner without
dispersing in time. This long-lived oscillation is a manifestation
of a permanent and periodic energy transfer between the axial and
transverse polarizations, which we consider as internal mode of the
vector solitary wave. To the best of our knowledge, this is the first
report of such oscillating solitary waves in continuum elastodynamics. 

Using initial conditions that generate a single solitary wave in each
numerical experiment, we have studied the velocity-amplitude relation,
and found that both types propagate faster at higher amplitudes. This
observation agrees with the result of \citet{ziv2019b}, who generated
a train of solitary waves of the first type, and found that the taller
waves in the train are faster than the shorter waves. Notably, the
vector solitary waves observed in the discrete mechanical systems
of \citet{deng2017,deng2019prl} exhibit an opposite behavior, namely,
they are faster at smaller amplitudes. Since the systems considered
by \citet{deng2017,deng2019prl} and the system considered here are
fundamentally different, this difference in the amplitude-dependent
nature of the vector solitary waves is reasonable. In the limit of
small strains, we have calculated velocity of low-frequency, long-wavelength
Bloch-Floquet waves in the laminate, and found it serves as a lower
bound the the velocity of the vector solitary waves. 

In the last part of the paper, we have described a procedure for selecting
initial strains that generate selected vector solitary waves.  We
emphasize that without the observation that the $\shearstrain-\pressurestrain$
relations of the laminate approximately follow their counterparts
in the homogeneous case, it would have not been possible to estimate
\emph{a priori} what should be the localized strains that generate
a certain solitary wave. We find this agreement remarkable, since
it correlates between two seemingly unrelated problems\textemdash impact
of homogeneous media and localized initial strain in laminated media,
and note it relies on the fact that Eq.$\ $\eqref{integralCurves}
is independent of the mass density. The procedure for generating selected
solitary waves also provides insights on the way that the axial strain
in the laminate changes as the resultant waves propagate. This allows
for an additional classification of the waves to tensile and compressive
solitary waves, according to their tendency to increase or decrease
that strain. 

Future work can follow different paths from this point. One possibility
is to attempt developing a homogenized system of equations \citep{yong2003solitary},
based on which it may be possible to carry out further analytical
investigation, \emph{e.g.}, as done by \citet{andrianov2013dynamic}
in the one-dimensional problem. This may be turn useful, for example,
in identifying the solution function that describes the quasi-shear
waves. Another possibility is to explore if additional types of vector
solitary waves will form when using different parameters for the compressible
Gent model than the set used here, or alternatively employ other types
of constitutive models, such as the \citet{ogden1972} and \citet{arru&boyc93jmps}
models. It would be also interesting to study the collision of vector
solitary waves in the laminate, which may result with anomalous phenomena,
as observed by \citet{deng2019prl} in their case of the discrete
two-dimensional mechanical model. As pointed out earlier, the vector
solitary waves reported here are different from those observed by
\citet{deng2019prl}, and hence expect that their response to collisions
will also be different. 

\section*{Acknowledgments}

We thank an anonymous reviewer whose constructive comments helped
improving this paper. We acknowledge the support of the Israel Science
Foundation, funded by the Israel Academy of Sciences and Humanities
(Grant no.$\ $1912/15), the United States-Israel Binational Science
Foundation (Grant no.~2014358), and Ministry of Science and Technology
(grant no.~880011).

\appendix


\section*{}

\bibliographystyle{natbib}

\end{document}